\documentclass[12pt]{article}  
\newcommand{\comma}{\;\; ,}
\newcommand{\period}{\;\; .}
\newcommand{\eq}{\; = \;}
\newcommand{\sep}{\;\; , \;\;}
\newcommand{\be}{\begin{equation}}
\newcommand{\bd}{\begin{displaymath}}
\newcommand{\ee}{\end{equation}}
\newcommand{\ed}{\end{displaymath}}
\newcommand{\ba}{\begin{eqnarray}}
\newcommand{\ea}{\end{eqnarray}}
\newcommand{\Wb}{\overline{W}}
\newcommand{\minus}{\! - \!}
\newcommand{\mod}{{\rm mod}\,}

\renewcommand{\i}{{\rm i}}
\newcommand{\e}{{\rm e}}

\newcommand{\er}{{\epsilon (r) }}
\newcommand{\inn}{\; {\scriptstyle \in } \;}
\newcommand{\notinn}{\; {\scriptstyle \notin } \;}

\newcounter{storeeqn}

\title{The order parameter of the chiral Potts model}

\author{ R.J. Baxter\\
{\protect \small  Mathematical
Sciences Institute}\\
{\protect  \small The Australian National University,
 Canberra, A.C.T. 0200,
  Australia, \small e-mail: none }}

\date{}

\begin{document}


\maketitle

\abstract{An outstanding problem in statistical mechanics is the 
order parameter of the chiral Potts model. An elegant  conjecture for 
this was made in 1983. It has since been successfully tested against 
series expansions, but there is as yet no
proof of the conjecture. Here we show that if one makes a certain 
analyticity assumption similar to that used to derive the free energy,
then one can indeed verify the conjecture. The method is based on the 
``broken rapidity line'' approach pioneered by Jimbo, Miwa and 
Nakayashiki.}


\vspace{5mm}

{{\bf KEY WORDS: } Statistical mechanics, lattice models, 
order parameter.}



\section{Introduction}

The chiral Potts model was originally formulated as an $N$-state
one-dimensional quantum hamiltonian \cite{HKN83, GR85}, and then 
as a two-dimensional classical lattice model in statistical 
mechanics.\cite{AYMPT87, MPT87, BPAuY88} It satisfies the star-triangle 
relations. The free energy was first obtained for the infinite 
lattice using the invariance properties of the free energy and its
derivatives.\cite{RJB88} Then in 1990 the functional transfer matrix 
relations of Bazhanov and Stroganov \cite{BazStrog90} were used to 
calculate  the free energy more explicitly as a double 
integral.\cite{BBP90, RJB90, RJB91}

The chiral Potts model is a system of spins living on the sites of a 
planar lattice, usually (but not necessarily \cite{RJB93}) taken to be 
the square lattice $\cal L$. Each spin takes one of $N$ possible states, 
labelled $0, \ldots , N-1$ and interacts with its neighbours. If the 
spin at site $i$ is $\sigma_i$, then the interaction between  
spins on adjacent sites $i$ and $j$ depends on $\sigma_i$, $\sigma_j$ 
only via their difference $\sigma_i - \sigma_j$ (mod $N$).

We shall define the interactions in the next section. Here we merely 
note that they depend on a temperature-like parameter $k'$, which is 
small at low temperatures and large at high temperatures. For $k' <1$,
a related  parameter is
\be \label{kkp}
k =  \surd (1- {k'}^2 )  \period \ee

Let $\omega = \e^{2 \pi \i/N}$ and let $a$ be a spin 
deep inside the lattice. Define 
\be {\cal M}_r \eq \langle \omega^{r a} \rangle  \ee
as the average value of $\omega^{r a}$, for $r = 0, \ldots , N$.
Suppose one fixes the boundary spins to be zero and allows the 
lattice to become infinitely large, $a$ remaining near the centre. 
Then for sufficiently high 
temperatures (weak enough interactions) the boundary conditions
will become irrelevant, all values of $a$ will be equally likely, 
and  ${\cal M}_r$ will be exactly zero for $r= 1, \ldots , N-1$.

However, the system displays ferromagnetic order. There is a 
{\em critical} temperature $T_c$  below which the boundary conditions
remain relevant even for an infinitely large lattice. Then 
${\cal M}_1, \ldots , {\cal M}_{N-1}$ are non-zero. They can be 
thought 
of as  spontaneous magnetizations or order parameters. We expect 
${\cal M}_r$ to  vanish as $T \rightarrow T_c$, being then proportional 
to
\bd 
(1-T/T_c)^{\beta}  \comma \ed
the index $\beta$ being known as a critical exponent (dependent 
on $r$). 

In fact the critical point is when $k'=1$ and $k = 0$, so the 
ferromagnetic region is when $0 < k,k' < 1$. 
For $N = 2$ the 
chiral Potts model reduces to the Ising model. {From} the exact 
results of Yang \cite{Yang52} 
and Onsager\cite{Onsager71}  we know that then ${\cal M}_1 = k^{1/4}$. 
Hence  ${\cal M}_1 = (1-{k'}^2)^{1/8}$ and the critical exponent 
$\beta$  is $1/8$.


In 1983 Howes, Kadanoff and den Nijs \cite{HKN83} considered 
the case $N=3$ and evaluated ${\cal M}_1 , {\cal M}_2$ to 
order ${k'}^{13}$ in a 
series expansion in powers of $k'$. They found their series
fitted the formula ${\cal M}_1 = {\cal M}_2 =  k^{2/9}$, 
giving $\beta = 1/9$.  Later, Henkel and  Lacki \cite{HK85} 
expanded  $\sum {\cal M}_r$ for general $N$ to order ${k'}^6$.

In 1989 Albertini {\it et al} \cite{AMPT89}  
expanded the general  $N$ case to  order ${k'}^5$, and found 
that all the results were 
consistent with the remarkably simple and elegant conjecture
\be \label{conj}
 {\cal M}_r \eq k^{r(N-r)/N^2} \; \; , \; \;  0 \leq r \leq N    
 \ee
(see also Ref. \cite{HK89}), giving $\beta = r(N-r)/2N^2$. Baxter 
\cite{RJB93} used finite corner transfer matrices to expand 
${\cal M}_r$  for $N=3$ to order ${k'}^{14}$, and again found the 
results were  consistent with the conjecture (\ref{conj}).

So for the last fifteen years or more there have been conjectures 
for the order parameters of the chiral Potts model.  However, the 
author knows of no derivation of any of these for $N > 2$.  One 
can contrast this with the situation for the Ising
model, where it was five years from the time Onsager calculated 
the free energy to when at a conference in Florence in 1949
he announced the formula for ${\cal M}_1$, and three years from 
then until Yang published a proof. (The actual calculation took
Yang six months of work off and on \cite[p. 12]{Yang83}.)

The difficulty has been that unlike most other solvable models, 
the chiral Potts model does 
{\em not} have the ``rapidity difference property''. For this reason 
the infinite-size ``corner transfer method'' \cite{book82} for exactly 
calculating the order parameters fails. An alternative method
was developed in 1993 by Jimbo {\it et al} \cite{JMN93}. It relies
on breaking one of the rapidity lines of the lattice, giving one 
half-line a different rapidity value from the other. The author 
applied this method to the chiral Potts in 1998 \cite{RJB98}
and wrote down functional relations for the resulting generalized 
order parameters $G_{pq}(r)$. They are functions of the rapidities 
$p, q$ of the two half-lines.

Again there was a difficulty. The relations by themselves do not 
completely define the $G_{pq}(r)$. 
One also needs information as to the analyticity
properties of the functions.\footnote{The example I like to quote 
is the relation $f(z+1) = f(z)$. By itself this merely says that 
$f(z)$ is periodic. However, if one can also show that  
$f(z)$ is analytic and bounded
in the strip  $ 0 \leq \Re (z) \leq 1$, then it follows that $f(z)$
is analytic everywhere and bounded, so by Liouville's theorem 
it is a constant.} In this respect the functional relations 
are similar to the ``inversion relations'' for the infinite-lattice
free energy \cite{RJB82, RJB03}.

The calculation of the free energy of the chiral Potts model
\cite{RJB90, RJB91, RJB03}
proceeds in two stages. First one considers a related  
``$\tau_2 (t_p)$'' model.\cite{RJB04} This is intimately 
connected with the superintegrable case of the chiral Potts 
model.\cite{RJB89} It is much simpler than the chiral Potts model
in that its Boltzmann weights depend on the horizontal
rapidity $p$ only via a single parameter $t_p$, and are linear in 
$t_p$. Its row-to-row transfer matrix is the product 
of two chiral Potts transfer matrices, one with horizontal 
rapidity $p$, the other with a related rapidity $\overline{p}(0,2)$
defined in eqn. (\ref{pkl}) of section 3. 

For a finite lattice, the partition function $Z$ of the 
$\tau_2 (t_p)$ model is therefore a polynomial in $t_p$.
The free energy is the logarithm of $Z^{1/M}$, where $M$ is the 
number of sites of the lattice, evaluated in the thermodynamic 
limit when the lattice becomes infinitely big. This limiting
function of  course may have singularities in the complex $t_p$
plane.  {\it A  priori}, one might expect it to have $N$ branch 
cuts, each running though one of the $N$ roots of unity. 
However, one can argue that in fact it only has one such cut. As 
a result the free energy (i.e. the maximum eigenvalue of the
transfer matrix) can be calculated by a Wiener-Hopf factorization.

 The second stage is to factor this free energy to obtain 
that of the chiral  Potts model.

It turns out that the first stage of this process can be applied to
the generalized order parameter function $G_{pq}(r)$, provided we
take $q$, $p$ to be related by eqn. (\ref{spcase}) of section 6.

We present the working in the following sections. We do not
need to calculate $G_{pq}(r)$ for general $p$, $q$ and we do not do so.

We define the model in section 2, and the function $G_{pq}(r)$ in
section 4. We also present the functional relations 
satisfied by $G_{pq}(r)$, but in fact we hardly use them. It is the 
analyticity properties that hold the key to calculating the $G_{pq}(r)$,
and most of the discussion in sections 3 to 6 is concerned
with presenting evidence for our assumptions regarding these properties.

In particular, we show that when $p, q$ satisfy  (\ref{spcase}),
the $G_{pq}(r)$ can be expressed in terms 
of $\tau_2 (t_p)$ Boltzmann weights $U(a,b,c,d)$ that are linear in 
$t_p$. 
We argue that $G_{pq}(r)$ is therefore like the free energy
of the  $\tau_2 (t_p)$ model, in that it has at most one 
branch cut in the $t_p$-plane, rather than the $N$ cuts that 
one might expect.  

We give our precise assumptions in section 7. Then in
section 8 we use them to obtain $G_{pq}(r)$ by a Wiener-Hopf 
factorization in very much the same way as one calculates the 
$\tau_2 (t_p)$ free energy. The desired formula 
(\ref{conj}) follows immediately. We also present an alternative 
method, looking at the product of $N$ such functions $G_{pq}(r)$,
that avoids the need for a Wiener-Hopf  factorization.

In section 9 we briefly discuss some other special cases (analogues of
the $\tau_j (t_p)$ models for  $j = 3, \ldots , N$) that may be 
tractable, and make a conjecture for the form of $G_{pq}(r)$ for
some of these cases.



\section{Chiral Potts model}

We use the notation of \cite{BPAuY88, BBP90, RJB98}. Let $k, k'$
be two real variables in the range $(0,1)$, satisfying (\ref{kkp}).
Consider four parameters
$a_p, b_p, c_p, d_p$ satisfying the homogeneous relations
\ba \label{abcd}
a_p^N + k' b_p^N  & = &  k d_p^N \comma \nonumber \\
k' a_p^N + b_p^N  & = &  k c_p^N \period \ea
Let $p$ denote the set $\{a_p, b_p, c_p, d_p \}$, or rather their 
ratios, since all the equations we shall write involve  
$a_p, b_p, c_p, d_p$ only via their ratios. Similarly $q$ denotes 
the set $\{a_q, b_q, c_q, d_q \}$ satisfying the relations 
(\ref{abcd}) (with $p$ replaced by $q$). We call $p$ and $q$
``rapidity'' variables.

Define Boltzmann weight functions $W_{pq}(n), \Wb _{pq}(n)$ by
\ba \label{WWb}
W_{pq}(n) & = & \prod_{j=1}^n \frac{d_p b_q - a_p c_q \omega^j}
{b_p d_q - c_p a_q \omega^j} \comma  \nonumber \\
\Wb_{pq}(n) & = & \prod_{j=1}^n \frac{\omega a_p d_q - d_p a_q  
\omega^j} {c_p b_q - b_p c_q \omega^j} \comma \ea
where
\bd \omega \eq \e^{2\pi \i/N} \period \ed

Then the conditions (\ref{abcd}) ensure that
\bd 
W_{pq}(n + N) = W_{pq}(n) \sep \Wb_{pq}(n + N) = \Wb_{pq}(n) 
\comma \ed
so the functions are periodic of period $N$. Note that
\bd 
W_{pq}(0 ) = \Wb _{pq}(0 ) =  1 \period \ed

Define related parameters
\be \label{xyt}
x_p = a_p/d_p \sep y_p = b_p/c_p \sep t_p = x_p y_p \sep
 \mu_p = d_p/c_p  \period \ee
They satisfy
\be \label{xymu} x_p^N+y_p^N = k(1+x_p^N y_p^N) 
\sep k x_p^N = 1-k'/\mu_p^N 
\sep k y_p^N = 1-k'\mu_p^N \comma \ee
\bd  k^2 t_p^N \eq  (1-k'\mu_p^N)(1-k'/\mu_p^N) \period \ed

We can also write (\ref{WWb}) as
\addtocounter{equation}{1}
\setcounter{storeeqn}{\value{equation}}
\setcounter{equation}{0}
\renewcommand{\theequation}{\arabic{storeeqn}\alph{equation}}
\ba \label{WWba}
W_{pq}(n) & = & (\mu_p/\mu_q)^n \prod_{j=1}^n \frac{y_q - \omega^j x_p}
{y_p - \omega^j x_q} \comma  \\
\label{WWbb}
\Wb_{pq}(n) & = &  (\mu_p \mu_q)^n  \prod_{j=1}^n \frac{\omega x_p -   
\omega^j x_q} {y_q - \omega^j y_p} \period \ea

\setcounter{equation}{\value{storeeqn}}
\renewcommand{\theequation}{\arabic{equation}}



\setlength{\unitlength}{1pt}
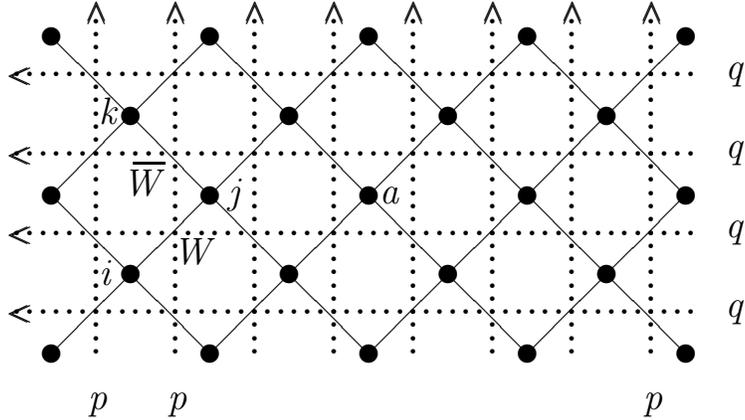
\begin{figure}[hbt]
\begin{picture}(420,160) (0,0)
\put (60,0) {\line(1,1) {120}}
\put (60,60) {\line(1,1) {60}}
\put (120,0) {\line(1,1) {120}}
\put (180,0) {\line(1,1) {120}}
\put (240,0) {\line(1,1) {60}}
\put (60,120) {\line(1,-1) {120}}
\put (60,60) {\line(1,-1) {60}}
\put (120,120) {\line(1,-1) {120}}
\put (180,120) {\line(1,-1) {120}}
\put (240,120) {\line(1,-1) {60}}
\multiput(60,0)(60,0){5}{\circle*{7}}
\multiput(60,60)(60,0){5}{\circle*{7}}
\multiput(60,120)(60,0){5}{\circle*{7}}
\multiput(90,30)(60,0){4}{\circle*{7}}
\multiput(90,90)(60,0){4}{\circle*{7}}

\multiput(45,15)(5,0){52}{\bf .}
\multiput(45,45)(5,0){52}{\bf .}
\multiput(45,75)(5,0){52}{\bf .}
\multiput(45,105)(5,0){52}{\bf .}

\thicklines

\multiput(73,125) (30,0){8}{{$\wedge$}}
\multiput(73,126) (30,0){8}{{$\wedge$}}

\multiput (39,12) (0,30){4}{ $< $}
\multiput (40,12) (0,30){4}{ $< $}

\thinlines

\multiput(75,0)(0,5){26}{\bf .}
\multiput(105,0)(0,5){26}{\bf .}
\multiput(135,0)(0,5){26}{\bf .}
\multiput(165,0)(0,5){26}{\bf .}
\multiput(195,0)(0,5){26}{\bf .}
\multiput(225,0)(0,5){26}{\bf .}
\multiput(255,0)(0,5){26}{\bf .}
\multiput(285,0)(0,5){26}{\bf .}
\put (315,14) {\large \it q }
\put (315,44) {\large \it q}
\put (315,74) {\large \it q}
\put (315,104) {\large \it q}
\put (74,-21) {\large \it p }
\put (104,-21) {\large \it p }
\put (284,-21) {\large \it p }
\put (78,26) {\large \it i}
\put (78,87) {\large \it k}
\put (127,57) {\large \it j}
\put (184,57) {\large \it a }
\put (106,34) {\large \it W }
\put (87,60) {\large \it {W} }
\put (91,72) {\line(1,0) {12}}
 \end{picture}
\vspace{1.5cm}
 \caption{\footnotesize The square lattice $\cal L$  of solid lines and 
circles,  showing the central spin $a$ and the dotted directed rapidity 
lines .}
 \label{sqlattice}
\end{figure}

Now consider the square lattice $\cal L$, drawn diagonally as in 
Figure \ref{sqlattice}, with a total of $M$ sites. On each site 
$i$ place a  spin $\sigma_i$, which can take any one of the $N$ 
values $0, 1, \ldots, N-1$. With each SW - NE edge $(i,j)$
(with $i$ below $j$) associate an edge weight 
$W_{pq}(\sigma_i - \sigma_j)$. 
Similarly, with each SW - NE edge $(j,k)$  ($j$ below $k$), associate
an edge weight  $\Wb_{pq}(\sigma_j - \sigma_k)$. Then the partition 
function is
\be Z \eq \sum_{\sigma} \, \prod W_{pq}(\sigma_i - \sigma_j)  \prod 
\Wb_{pq}(\sigma_j - \sigma_k) \comma \ee
the products being over all edges of each type, and the sum over all 
$N^M$ values of the $M$ spins. We expect the partition function 
per site
\bd \kappa \eq Z^{1/M} \ed
to tend to a unique limit as the lattice becomes large in both 
directions.

Let $a$ be a spin on a site near the centre of the lattice, as in 
Figure \ref{sqlattice},  and let $f(a)$ be any function thereof. 
Then the thermodynamic average  of   $f(a)$ is 
\be \label{avfa}
\langle f(a) \rangle \eq Z^{-1} \, \sum_{\sigma} \, f(a) 
 \prod W_{pq}(\sigma_i - \sigma_j)  \prod  
\Wb_{pq}(\sigma_j - \sigma_k) \period \ee
we expect this to also tend to a limit as the lattice
becomes large.

We have also shown in Figure \ref{sqlattice} the vertical
and horizontal ``rapidity lines''. Each edge of $\cal L$  passes 
through the intersection of two rapidity lines.

We could allow each vertical (horizontal) rapidity line $\alpha$ to 
have a different rapidity $p_{\alpha}$ ($q_{\alpha}$). If an edge 
of $\cal L$ lies on lines with rapidities $p_{\alpha}$, $q_{\beta}$,
then the 
Boltzmann weight function of that edge 
is to be taken as $W_{pq}(n)$ or $\Wb_{pq}(n)$, with
$p = p_{\alpha}$ and $q = q_{\beta}$.

The weight functions $W_{pq}(n)$,  $\Wb_{pq}(n)$ satisfy the star-
triangle relation.\cite{BPAuY88} For this reason we are free to move
the rapidity lines around in the plane, in particular to interchange 
two vertical or two horizontal rapidity lines.\cite{RJB78} So long 
as no rapidity 
line crosses the site  with spin $a$ while making such rearrangements, 
the  average $\langle f(a) \rangle$ is {\em unchanged} by the  
rearrangement.\footnote{Subject to boundary conditions: here we are 
primarily interested in
the infinite lattice, where we expect the  boundary conditions
to have no effect on the rearrangements 
we consider.}

All of the rapidity lines shown in Figure \ref{sqlattice} are
``full'', in the sense that they extend without break from
one boundary to another. We can move any such 
line away from the central site
to infinity, where we do not expect it to contribute to 
$\langle f(a) \rangle$. Hence
$\langle f(a) \rangle$ {\em must be independent of all the
full-line $p$ and $q$ rapidities}.

It can still depend on $k$ or $k'$, which play the role of 
universal constants, the same for all sites of $\cal L$. As we 
mentioned in the introduction, it has been conjectured that
\be \label{conj2}
{\cal M}_r \eq \langle \omega^{r a} \rangle \eq 
k^{r(N-r)/N^2} \; \; , \; \;  0 \leq r \leq N    \period \ee
The aim of this paper is to verify this conjecture, subject
to a plausible  analyticity assumption.




\section{$\tau_2(t_p)$ model}

One of the difficulties of the chiral Potts model is the
multi-valued nature of the relations between $x_p, y_p , \mu_p$.
Every such relation involves taking an $N$th root. For $N = 2$
the model reduces to the Ising model. In this case there is a simple
uniformizing substitution whereby all variables can be written as 
single-valued Jacobi elliptic functions of another variable 
$u_p$.\cite[App. B]{RJB03} For higher values of $N$ no such 
substitution is known. It is therefore significant that the chiral 
Potts model can be related to ``superintegrable'' or
 ``$\tau_j(t_p)$''
models, where the dependence of Boltzmann weights on the horizontal 
rapidity $p$ is simple: they are explicit polynomials in
the single variable $t_p$. (They still involve all the vertical 
rapidity variables $x_v, y_v , \mu_v$.)

The key equations are given in \cite{BBP90}. Let us change notation
and use the symbols $v$ or $v'$ for the vertical rapidities, $p$, 
$q$ for the rapidities of the lower and upper rows. Thus
we replace the $p, p', q, r$ of \cite{BBP90} by $v,v',p,q$.

For generality, we allow $v$ and $v'$ to de different in this section.
In later sections we take $v'= v$ and all vertical rapidities to 
be the same. $p$ and $q$ will play the role of variables in the 
functional relations we discuss, while we shall regard $v$ as a 
constant. 

Consider 
two rows of 
edges of $\cal L$. Let the horizontal rapidity of the lower row be 
$p$, and of the upper row 
\be \label{pkl}
q \eq \overline{p}(k,\ell) \comma \ee 
by which we mean
\be \label{rowreln}
x_{q} = \omega^k \, y_{p} \sep y_{q} = \omega^{\ell} x_{p}
\sep \mu_{q} = 1/\mu_{p} \comma \ee
$k, \ell$ being integers. We can impose the restriction
\be 1 \leq k+\ell \leq N \period \ee

If we sum over an intermediate spin between these rows, we 
construct a combined weight function 
\be \label{Uwt}
U(a,b,c,d) \eq \sum_{g = 0}^{N-1}  W_{v p}(a \minus g) \, 
\Wb_{v \, q}(g \minus d)
\, \Wb_{v' p}(b \minus g) \, W_{v' q}(g \minus c) \comma \ee
depicted in Figure \ref{Uabcd} This can in turn be written as
\be \label{UVV} 
U(a,b,c,d) \eq N \, \sum_{n=0}^{N-1} V_{v p q}(a,d;n) \,  
V_{v' q p}(-c,-b;n) \comma \ee
where
\be \label{defVvpq}
V_{v p q}(a,d;n) \eq N^{-1} \, \sum_{g=0}^{N-1} \omega^{n g} \,
 W_{v p}(a-g) \,  \Wb_{v q}(g-d) \period \ee



\setlength{\unitlength}{1pt}
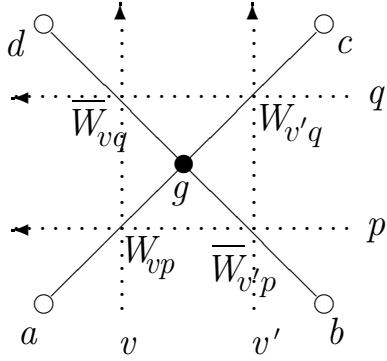
\begin{figure}[hbt]
\begin{picture}(420,160) (0,0)
\put (160,0) {\line(1,1) {100}}
\put (160,100) {\line(1,-1) {100}}

\put (210,50) {\circle*{7}}
\put (157,-3) {\circle{7}}
\put (263,-3) {\circle{7}}
\put (157,103) {\circle{7}}
\put (263,103) {\circle{7}}

\multiput(145,25)(5,0){26}{.}
\multiput(145,75)(5,0){26}{.}

\thicklines
\put (146,25) {\vector(-1,0) {1}}
\put (146,75) {\vector(-1,0) {1}}
\put (186,110) {\vector(0,1) {1}}
\put (236,110) {\vector(0,1) {1}}
\thinlines

\multiput(185,-5)(0,5){23}{.}
\multiput(235,-5)(0,5){23}{.}
\put (279,23) {\large \it p}
\put (279,73) {\large \it q}

\put (185,-22) {\large \it v}
\put (235,-22) {\large \it v}
\put (244,-21) {$'$}
\put (147,-18) {\large \it a}
\put (205,37) {\large \it g}
\put (267,92) {\large \it c}
\put (264,-18) {\large \it b}
\put (142,91) {\large \it d}
\put (185,13) {\large \it W}
\put (195,9) {\it vp}

\put (236,63) {\large \it W}
\put (248,59) {$v'q$}

\put (221,19) {\line(1,0) {12}}
\put (218,6) {\large \it W}
\put (230,2) {$v'p$}

\put (168,74) {\line(1,0) {12}}
\put (165,61) {\large \it W}
\put (177,57) {$vq$}
 \end{picture}
\vspace{1.5cm}
 \caption{\footnotesize The combined weight $U(a,b,c,d)$ of four edges, 
summed over the middle spin $g$.}
 \label{Uabcd}
\end{figure}

Let $\zeta_{k \ell}$ be the set of integers 
$\{ -k,1-k,\ldots, \ell-1\}$,
with $k+\ell$ elements. We say that $n \inn \zeta_{k \ell}$ if
$n$ is equal, modulo $N$, to one of the elements of $\zeta_{k \ell}$.
Then in (3.16), (3.17) of \cite{BBP90} it is shown that
\ba  V_{v p q}(a,d;n) & = & 0 \; \; {\rm if} \; \; a-d \inn 
\zeta_{k \ell}
\; \; {\rm and } \; \; n \notinn \zeta_{k \ell} \; , \nonumber \\
 V_{v q p}(-c,-b;n) & = & 0 \; \; {\rm if} \; \; n \inn \zeta_{k \ell}
\; \; {\rm and } \; \; b-c \notinn \zeta_{k \ell} \; ,  \\
U(a,b,c,d) & = & 0 \; \; {\rm if} \; \; a-d \inn \zeta_{k \ell}
\; \; {\rm and } \; \; b-c \notinn \zeta_{k \ell} \; \; . \nonumber 
\ea


\setlength{\unitlength}{1pt}
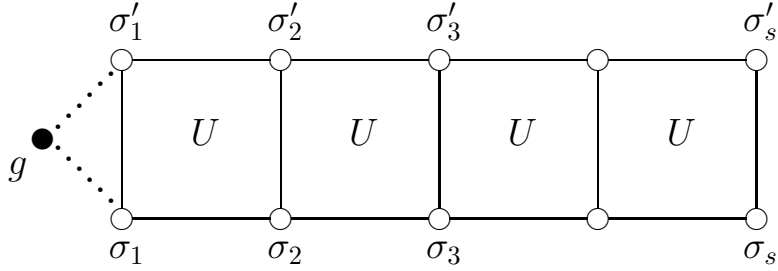
\begin{figure}[hbt]
\begin{picture}(420,160) (-20,0)
\multiput(80,4)(60,0){5}{\line(0,1) {52}}
\multiput(84,0)(60,0){4}{\line(1,0) {52}}
\multiput(84,60)(60,0){4}{\line(1,0) {52}}
\put (50,30) {\circle*{8}}
\put(54,34) {\bf .}
\put(58,38) {\bf .}
\put(62,42) {\bf .}
\put(66,46) {\bf .}
\put(70,50) {\bf .}
\put(74,54) {\bf .}
\put(54,26) {\bf .}
\put(58,22) {\bf .}
\put(62,18) {\bf .}
\put(66,14) {\bf .}
\put(70,10) {\bf .}
\put(74,6) {\bf .}
\multiput(80,0)(60,0){5}{\circle{8}}
\multiput(80,60)(60,0){5}{\circle{8}}
\put (75,-15) {\large $\sigma_1$}
\put (135,-15) {\large $\sigma_2$}
\put (195,-15) {\large $\sigma_3$}
\put (315,-15) {\large $\sigma_s$}
\put (75,71) {\large $\sigma'_1$}
\put (135,71) {\large $\sigma'_2$}
\put (195,71) {\large $\sigma'_3$}
\put (315,71) {\large $\sigma'_s$}
\put (106,28) {\large $U$}
\put (166,28) {\large $U$}
\put (226,28) {\large $U$}
\put (286,28) {\large $U$}
\put (37,17) {\large $g$}
 \end{picture}
\vspace{1.5cm}
 \caption{\footnotesize Two rows of spins with face weight functions 
$U(\sigma_i, \sigma_{i+1}, \sigma'_{i+1}, \sigma'_i)$
in between, and a spin $g$ to the left linked by edge weights to 
$\sigma_1, \sigma'_1$.}
 \label{tworows}
\end{figure}

Consider a row of adjacent spins $\sigma_1, \ldots , 
\sigma_s$,  below a similar row $\sigma'_1, \ldots , \sigma'_s$,
as in Figure \ref{tworows}, with intervening face
weights   (\ref{Uwt}). For the moment ignore the spin $g$ and the 
dotted lines. The combined weight of the rows is
\be \label{combwt}
I( \sigma_1, \ldots ,\sigma'_s ) \eq  \prod_{i=1}^{s-1} 
U(\sigma_i, \sigma_{i+1}, \sigma'_{i+1}, 
\sigma'_i)  \ee
and it follows that if $\sigma_1 - \sigma'_1 \inn \zeta_{k \ell}$,
then the weight is zero unless $\sigma_i - \sigma'_i \inn 
\zeta_{k \ell}$ for $i = 1, \ldots, s$. In this case
we only need $U(a,b,c,d)$ for $a-d$ and $b-c$ in $\zeta_{k \ell}$.
{From} (3.21), (3.39) of \cite{BBP90}, if we define
\bd  j = k+\ell \sep \alpha = \mod(a-d+k,N) \ed 
\be  m = \mod(n+k,N) \sep 
\beta = \mod(b-c+k,N) \; , \ee
then

\be \label{Veqns}
V_{v p q} (a,d;n) \eq \Omega_{v p}^{kl} \, \omega^{nd-mk}
\, (b_p/d_p)^{\alpha} y_p^{-m} f_{v p}(j,\alpha,m) \ee
\bd V_{v' q p} (-c,-b;n) \eq \overline{\Omega}_{v' p}^{kl} \, 
\overline{h}_{v' p}^{(j)} \, \omega^{k c - m b}
(d_p/b_p)^{\beta} \frac{\eta_{p,j,\beta}}{\eta_{p,j,m}}
 \, y_p^{m} f_{v'p}(j,\beta,m) \ed
provided $a-d, n,  b-c$ are all in $\zeta_{k \ell}$, which means 
that
\be 0 \leq \alpha, m, \beta <  j \leq N \period \ee

The expressions (\ref{Veqns}) become less frightening if one
groups them into factors of various types:


\subsubsection*{(i)  Factors independent of 
$a,b,c,d,n,\alpha, \beta, m$.}
These are the factors $\Omega_{v p}^{kl} $, 
$\overline{\Omega}_{v' p}^{kl}$, $\overline{h}_{v' p}^{(j)}$. They  
are defined in (3.24) and (3.35) of \cite{BBP90}. In this paper we
do not calculate full partition functions, but rather ratios of 
partition functions  such as (\ref{FZZ}) and (\ref{defGpq}).
These factors cancel out of such ratios, so we can ignore them.
Hereinafter we shall simply take $\Omega_{v p}^{kl}  = $ 
$\overline{\Omega}_{v' p}^{kl} = $ $\overline{h}_{v' p}^{(j)} = 1$
in (\ref{Veqns}).

\subsubsection*{(ii)  Factors that are powers of  $(b_d/d_p)$.}
These are the factors  $(b_p/d_p)^{\alpha}$, 
$(d_p/b_p)^{\beta}$. In Figure \ref{tworows} each internal
vertical edge $(\sigma_j, \sigma'_j)$ acquires a weight 
$(d_p/b_p)^{\beta}$ from the face on its left, and a factor
$(d_p/b_p)^{-\beta}$ from the face on its right, where
$\beta = \mod (\sigma_j- \sigma'_j +k,N)$. The contributions
to the internal edges therefore cancel.

\subsubsection*{(iii)  Factors  $y_p^{-m}, y_p^m$.}
These cancel from the product on the rhs of (\ref{Uwt}).

\subsubsection*{(iv)  The $\omega, \eta, f_{vp}$ factors.}
These remaining factors depend on the rapidity $p$ only via the 
functions $\eta_{p,j,\alpha}, f_{vp}(j,\alpha,m)$. These are defined 
in (3.26), (3.37) and (3.38) of \cite{BBP90}
(where our $f_{vp}$ becomes $F_{p q}$). They have a vital property:
{\em they are Laurent polynomials in $t_p$}, with no other dependence 
on $p$. More strongly, $\eta_{p,j,\alpha}$ is proportional to 
$t_p^{\alpha}$, and $f_{vp}(j,\alpha,m)$ is a polynomial in $t_p$
of degree not greater than $m$ and $j-\alpha-1$, with minimum power
not less than zero and  $m-\alpha$.

It follows from 
(\ref{UVV} ) that $(d_p/b_p)^{\alpha-\beta} \, U(a,b,c,d)$ is a 
polynomial in $t_p$ of degree $j-1$. Hence from 
 (\ref{combwt})
\be \label{I1}
I(\sigma_1, \ldots , \sigma'_s ) \eq (b_p/d_p)^{\lambda-\nu} \; 
{\cal I} (\sigma_1, \ldots \sigma'_s) \comma \ee
where ${\cal I} (\sigma_1, \ldots \sigma'_s)$ is a polynomial in $t_p$  
of degree $(s-1)(j-1)$ and $\lambda = \mod (\sigma_1 - \sigma'_1+k,N)$, 
$\nu = \mod (\sigma_s - \sigma'_s+k,N)$. 

A related quantity that we shall need is
\be \label{defJr}
J_r(\sigma_1, \ldots \sigma'_s) \eq  V_{v'qp}(-\sigma'_1,-\sigma_1;-r)
\, I(\sigma_1, \ldots \sigma'_s) \comma \ee
where $r$ is some integer. This corresponds to including the spin $g$ 
in Figure \ref{tworows}, together with weight functions between $g$ and 
$\sigma_1$, $\sigma'_1$, multiplying by $\omega^{r g}$, and summing 
over $g$.
The type (ii) factors now cancel from the 
left-hand edge $(\sigma_1, \sigma'_1)$, but a type (iii) factor 
$y_p^m$ arises 
from  the extra $ V_{v'qp}$ term. Using $y_p = t_p/x_p$, it follows 
that
\be \label{J1}
x_p^{m(r)} J_r(\sigma_1, \ldots \sigma'_s) \eq    (d_p/b_p)^{\nu} 
t_p^{j-1} \,  {\cal J}_r(\sigma_1, \ldots \sigma'_s) \comma \ee
where
$m(r) = \mod (k-r,N)$, $\nu = \mod (\sigma_s-\sigma'_s+k,N)$,
and  ${\cal J}(t_p) $ is a Laurent polynomial in $t_p$.


\subsubsection*{The case $j = 2$}
One case in which we shall be particularly interested  is when 
\be\label{jeq2}
 k=0 \sep \ell=2 \sep j=2  \period \ee
For this case the functions $\eta, f_{v p}(j, \alpha, m) $ are 
given in (3.48) of \cite{BBP90}. {From} (\ref{UVV}), 
\be \label{UUhat}
U(a,b,c,d) \eq \frac{N}{y_v y_{v'}} \, 
\left( \frac{b_p}{d_p} \right)^{a+d-b-c}
\widehat{U}(a,b,c,d) \ee
where $\widehat{U}(a,b,c,d)$ is given in Table \ref{Uwts}.


\begin{table}[hbt]
\begin{center}
\begin{tabular} {| c | c | c |}
\multicolumn{3}{c} {} \\
\hline
$ a \minus  d $  & $ b \minus c$  & $ \widehat{U}(a,b,c,d)$ \\ \hline
0  &  0 & $ y_v y_{v'} - \omega^{d-b+1} t_p  $ \\
0  &  1 & $ - \omega \mu_{v'} t_p (y_v  - \omega^{d-b+1} x_{v'} )  $ \\
1   &  0 & $  \mu_{v}  (y_{v'}  - \omega^{d-b+1} x_{v} )  
$ \\
1  &  1 & $ - \omega \mu_{v} \mu_{v'}  (t_p  - \omega^{d-b+1} x_{v} 
x_{v'} )  $ \\
\hline
\end{tabular}
\end{center}
\caption{\footnotesize The face weights  $\widehat{U}(a,b,c,d)$ of the
 $\tau_2(t_p)$ model.}
\vspace{0.5cm}
\label{Uwts}
\end{table}

In particular, from (\ref{combwt}) and (\ref{I1}) it follows that
\be {\cal I}(0, \ldots , 0) \eq (y_v y_{v'}- \omega^{d-b+1} t_p)^{s-1} 
\comma \ee
ignoring factors independent of $p$.

\subsubsection*{The case $j = N$}

The other case we shall need is when 
\be \label{3rdcase}
k=-1 \sep \ell = N+1  \sep j=N \period \ee

Then $\zeta_{k\ell}$ is the full set of $N$ integers $1, \ldots ,N$
and $V_{vpq}(a,d;n), V_{v,q,p}(-c,-b;n)$ are always given by 
(\ref{Veqns}), with $\alpha ,\beta , m$ in the range $[0,N-1]$.

In this case we shall be interested in the modified product
(\ref{defJr}), when
 $\sigma_s = \sigma'_s = 0$. Then $J_r(\sigma_1, \ldots , \sigma'_s)$
is given by  (\ref{J1}), with
$\nu = N-1$. The factor $(d_p/b_p)^{\nu}$ is now independent of 
$r$ and the spins $\sigma_1, \ldots, \sigma'_{s-1}$. Like the type (i)
factors, it cancels out of the ratios of interest  (\ref{FZZ}) 
and (\ref{defGpq}). The function 
${\cal J}_r (\sigma_1, \ldots \sigma'_s)$ in (\ref{J1}) is
a polynomial in $t_p$ of degree $(s-1)(N-1)$.

One particular sub-case that we shall consider is when
$\sigma_1 = \cdots = \sigma'_s = 0 $. {From} (3.26), (3.37), (3.38) of
\cite{BBP90} we can verify that 
\be
\eta_{p,N,\alpha} \eq  t_p^{\alpha} \period \ee
\be 
f_{vp}(N, N-1, m) \eq (d_v/b_v)^{N-1} x_v^m \comma \ee
and hence from (\ref{Uwt}) and (\ref{Veqns}) that
\be U(0,0,0,0) \eq \prod_{j=1}^{N-1} (x_v x_{v'}  - \omega^j t_p)^{s-1}
\comma \ee
ignoring  factors independent of $p$.
It follows that
\be \label{Jrlt}
{\cal J}_r(0, \ldots , 0) \eq 
 \prod_{j=1}^{N-1} (x_v x_{v'}  - \omega^j t_p)^{s-1} \period \ee




\section{The generalized order parameter}

Now let us replace all the vertical rapidities $p$ in Figure
\ref{sqlattice}
by $v$. We also replace all horizontal rapidities $q$ by $h$, except for
the one line immediately below the spin $a$. 
Following Jimbo, Miwa and Nakayashiki\cite{JMN93}, we break this line
immediately below the site $i$ containing the spin $a$, and give 
the half-line to the left the rapidity 
$p$, that to the right the rapidity $q$, as indicated in Figure
\ref{Fpq1}. Let $F_{pq}(a)$ be the probability that the spin at
site $i$ is in state $a$, i.e. from (\ref{avfa}),
\be  \label{FZZ}
 F_{pq}(a) \eq \langle \delta_{\sigma_i, a} \rangle  \eq Z(a)/Z 
\comma  \ee
where $Z(a)$ is the sum-over-states with spin $\sigma_i$ fixed to 
be $a$, divided by the unrestricted sum $Z = Z(0) +  \cdots + Z(N-1)$.



\setlength{\unitlength}{1pt}
\begin{figure}[hbt]

\begin{picture}(420,260) (0,0)

\multiput(30,15)(5,0){73}{.}
\multiput(30,75)(5,0){32}{\bf .}
\multiput(31,75)(5,0){32}{\bf .}
\multiput(202,75)(5,0){35}{\bf .}
\multiput(203,75)(5,0){35}{\bf .}
\multiput(30,135)(5,0){73}{.}
\multiput(30,195)(5,0){73}{.}
\put (190,72) {\line(0,1) {8}}
\put (200,72) {\line(0,1) {8}}
\thicklines

\put (69,72) {\large $< $}
\put (70,72) {\large $< $}
\put (71,72) {\large $< $}

\put (308,12) {\large $< $}
\put (309,12) {\large $< $}
\put (310,12) {\large $< $}

\put (308,72) {\large $< $}
\put (309,72) {\large $< $}
\put (310,72) {\large $< $}

\put (308,132) {\large $< $}
\put (309,132) {\large $< $}
\put (310,132) {\large $< $}

\put (308,192) {\large $< $}
\put (309,192) {\large $< $}
\put (310,192) {\large $< $}

\put (42,230) {\large $\wedge$}
\put (42,229) {\large $\wedge$}
\put (42,228) {\large $\wedge$}

\put (102,230) {\large $\wedge$}
\put (102,229) {\large $\wedge$}
\put (102,228) {\large $\wedge$}

\put (162,230) {\large $\wedge$}
\put (162,229) {\large $\wedge$}
\put (162,228) {\large $\wedge$}

\put (222,230) {\large $\wedge$}
\put (222,229) {\large $\wedge$}
\put (222,228) {\large $\wedge$}

\put (342,230) {\large $\wedge$}
\put (342,229) {\large $\wedge$}
\put (342,228) {\large $\wedge$}

\thinlines


\put (176,102) {{\Large \it a}}
\put (320,60) {{\Large \it q}}
\put (83,60) {{\Large \it p}}
\put (380,-2) {{\Large \it h}}
\put (380,118) {{\Large \it h}}
\put (380,178) {{\Large \it h}}

\put (195,105) {\circle{7}}

\put (16,45) {\line(1,-1) {60}}
\put (16,165) {\line(1,-1) {180}}
\put (76,225) {\line(1,-1) {117}}
\put (198,103) {\line(1,-1) {117}}
\put (196,225) {\line(1,-1) {180}}
\put (316,225) {\line(1,-1) {60}}
\put (16,165) {\line(1,1) {60}}
\put (16,45) {\line(1,1) {180}}
\put (76,-15) {\line(1,1) {117}}
\put (198,107) {\line(1,1) {118}}
\put (196,-15) {\line(1,1) {180}}
\put (316,-15) {\line(1,1) {60}}

\put (75,105) {\circle*{7}}
\put (315,105) {\circle*{7}}
\put (75,-15) {\circle*{7}}
\put (195,-15) {\circle*{7}}
\put (315,-15) {\circle*{7}}

\put (15,45) {\circle*{7}}
\put (135,45) {\circle*{7}}
\put (255,45) {\circle*{7}}
\put (375,45) {\circle*{7}}

\put (15,165) {\circle*{7}}
\put (135,165) {\circle*{7}}
\put (255,165) {\circle*{7}}
\put (375,165) {\circle*{7}}

\put (75,225) {\circle*{7}}
\put (195,225) {\circle*{7}}
\put (315,225) {\circle*{7}}

\put (42,-40) {{\Large \it v}}
\put (102,-40) {{\Large \it v}}
\put (162,-40) {{\Large \it v}}
\put (222,-40) {{\Large \it v}}
\put (282,-40) {{\Large \it v}}
\put (342,-40) {{\Large \it v}}

\multiput(45,-25)(0,5){52}{.}
\multiput(105,-25)(0,5){52}{.}
\multiput(165,-25)(0,5){52}{.}
\multiput(225,-25)(0,5){52}{.}
\multiput(285,-25)(0,5){52}{.}
\multiput(345,-25)(0,5){52}{.}
 \end{picture}

\vspace{1.5cm}
\caption{\footnotesize  First picture of the function $F_{pq}(a)$.}
\label{Fpq1}
\end{figure}
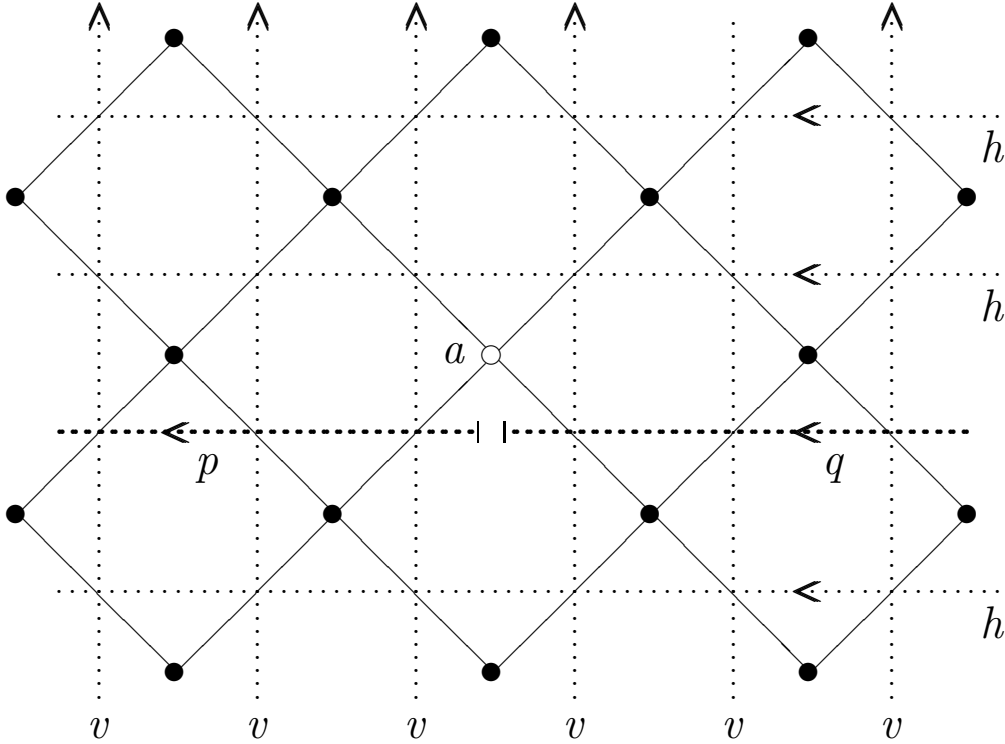

Because of the star-triangle relation, $F_{pq}(a)$ is independent
of the background rapidities $v$ and $h$, but it does depend on $p$ and 
$q$. This is because the ends of the  half-lines near $a$ cannot be 
moved away from $a$.

In \cite{RJB98} we show that one can obtain functional relations
satisfied by $F_{pq}(a)$. Let $R, S$ be the automorphisms defined in
\cite{BPAuY88}:
\be \label{Rop}
\{ a_{Rp},  b_{Rp},  c_{Rp},  d_{Rp} \} \eq \{ b_p, 
\omega a_p, d_p, c_p \} \comma \ee 
\be \{ a_{Sp},  b_{Sp},  c_{Sp},  d_{Sp} \} \eq \{\omega^{-1/2}  c_p, 
d_p, a_p, \omega^{-1/2} b_p \} \comma \ee 
so 
\be \label{xyR}
x_{Rp} = y_p \sep y_{Rp} =  \omega x_p \sep \mu_{Rp} = 
1/\mu_p \comma \ee
\be  \label{xyS}
x_{Sp} = 1/y_p \sep y_{Sp} = 1/ x_p \sep \mu_{Sp} = \omega^{-1/2}
y_p/(x_p \mu_p) \period \ee

Then from equations (18), (19) of \cite{BPAuY88},
\bd W_{pq}(n) = \Wb_{q,Rp}(-n) \sep \Wb_{pq}(n) = W_{q,Rp}(n) \ed
\be \label{RR}
W_{Rp, Rq}(n) = W_{pq}(-n) \sep \Wb_{Rp, Rq}(n) = \Wb_{pq}(-n) \ee
\bd  W_{Sp,Sq}(n) = W_{qp}(n) \sep  \Wb_{Sp,Sq}(n) = \Wb_{qp}(-n)  \ed
\be \label{SS}
W_{Sp, RSq}(n) = \Wb_{pq}(-n) \sep \Wb_{Sp, RSq}(n) = W_{pq}(-n) 
\period \ee

\subsubsection*{Symmetries}

{From} (\ref{RR}), operating by $R$ on all rapidities is equivalent to 
negating all  spins. Hence
\be \label{FRR} 
F_{Rp, Rq}(a) \eq F_{pq} (-a) \period \ee

There is a reflection symmetry that is not given in \cite{RJB98}. 
Operate  by $S$ on all the vertical rapidities, and by $RS$ on
all the horizontal rapidities in Figure \ref{Fpq1}. Then from 
(\ref{SS}) this is equivalent to interchanging the functions $W$, $\Wb$,
and negating all spins. Interchanging $W$ with $\Wb$ is in turn 
equivalent to mirror-reflecting the lattice about the central vertical 
line thorough $a$, while interchanging $p$ with $q$. Thus
\be \label{FRS}
F_{RSp, RSq}(a) \eq F_{qp}(-a) \period \ee

The dotted rapidity lines in Figures \ref{sqlattice} and \ref{Fpq1}
are directed, bearing arrows that give their direction. They form a 
graph $\cal G$ (the square lattice) of coordination number 4. The 
dual of $G$ is a bi-partite graph, and one of the two sub-graphs
is the lattice $\cal L$.

Although one cannot remove the ends of the half-lines from
the spin $a$, one can rotate them subject to the rules given in 
\cite{RJB98}, notably that one is not allowed to introduce any directed 
circuits into $\cal G$. The effect of this is to deform $\cal G$, but
the sites of  $\cal L$ continue to live on one of the two sub-lattices
dual to  $\cal G$. Every edge of $\cal L$ passes through the 
intersection of two dotted rapidity lines of $\cal G$. If the two 
arrows on 
$\cal G$ lie on either side of the edge , with rapidities $p,q$,
oriented as for the edge $(j,k)$ of Figure \ref{sqlattice}, then the 
weight function is  $\Wb_{pq}(k-j)$. If both arrows lie on one side 
of the edge, oriented as for the edge $(i,j)$, then the
weight function is  $W_{pq}(k-j)$. 

We show in Figure 8 of \cite{RJB98} that if the arrows of all 
rapidity lines crossing a given line of rapidity 
$p$ point to the left (right), one may reverse the arrow on $p$ and 
replace $p$ by $Rp$ ($R^{-1}p$).

The result is that one can perform the following
sequence of operations:

a) rotate the left half-line $p$ clockwise through  $90^{\circ}$ to a 
vertical  position below $a$, pointing downwards

b) replace $p$ by $R^{-1}p$ and reverse the arrow to point upwards,

c) rotate this half-line $R^{-1}p$ clockwise though another 
$90^{\circ}$ to the horizontal  position of Figure \ref{Fpq2}.


\setlength{\unitlength}{1pt}
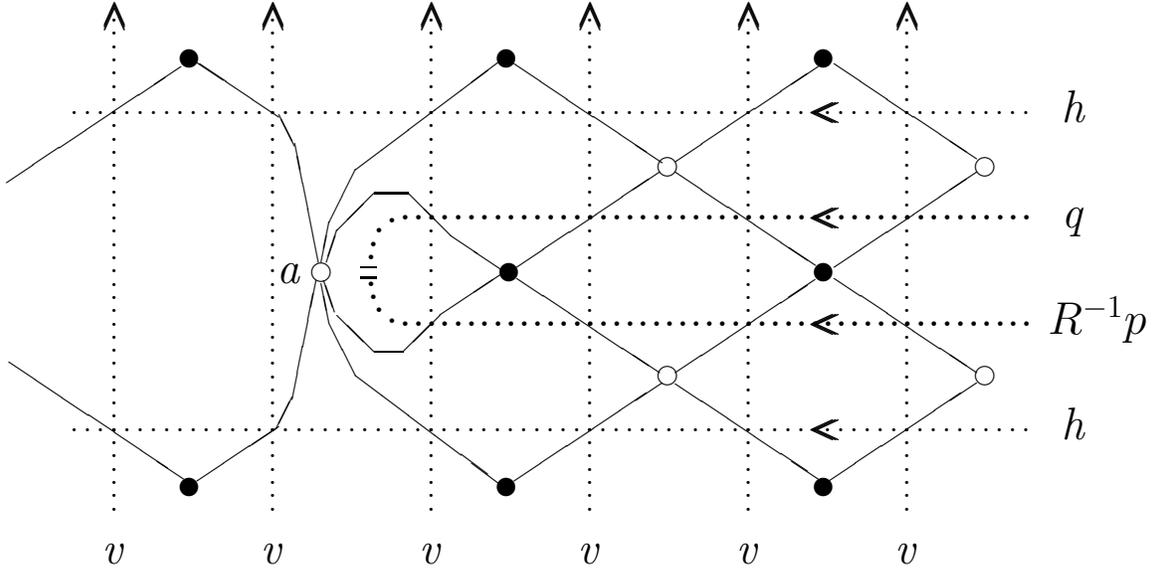
\begin{figure}[hbt]
\begin{picture}(420,260) (0,0)

\multiput(30,15)(5,0){73}{.}
\multiput(155,55)(5,0){48}{\bf .}
\multiput(155,95)(5,0){48}{\bf .}
\multiput(30,135)(5,0){73}{.}
\put (150,94) {\bf .}
\put (145,90) {\bf .}
\put (143,85) {\bf .}
\put (142,80) {\bf .}
\put (150,56) {\bf .}
\put (145,60) {\bf .}
\put (143,65) {\bf .}
\put (142,70) {\bf .}
\put (140,73) {\line(1,0) {6}}
\put (140,77) {\line(1,0) {6}}
\put (125,75) {\circle{7}}
\put (196,75) {\circle*{7}}
\put (75,-6) {\circle*{7}}
\put (75,156) {\circle*{7}}
\put (195,-6) {\circle*{7}}
\put (195,156) {\circle*{7}}
\put (256,36) {\circle{7}}
\put (256,115) {\circle{7}}
\put (315,-6) {\circle*{7}}
\put (315,75) {\circle*{7}}
\put (315,156) {\circle*{7}}
\put (376,36) {\circle{7}}
\put (376,115) {\circle{7}}


\put (124,79) {\line(-1,5) {9}}
\put (115,123) {\line(-1,2) {6}}
\put (109,134) {\line(-3,2) {31}}

\put (114,27) {\line(1,5) {9}}
\put (108,16) {\line(1,2) {6}}
\put (78,-4) {\line(3,2) {31}}

\put (125,79) {\line(1,5) {3}}
\put (128,94) {\line(1,2) {10}}
\put (138,114) {\line(4,3) {54}}

\put (127,79) {\line(1,3) {4}}
\put (131,91) {\line(1,1) {14}}
\put (145,105) {\line(1,0) {13}}
\put (158,105) {\line(1,-1) {16}}
\put (174,89) {\line(3,-2) {19}}

\put (125,71) {\line(1,-5) {3}}
\put (128,56) {\line(1,-2) {10}}
\put (138,36) {\line(4,-3) {54}}

\put (6,109) {\line(3,2) {66}}

\put (7,41) {\line(3,-2) {66}}

\put (126,72) {\line(1,-3) {4}}
\put (131,59) {\line(1,-1) {14}}
\put (145,45) {\line(1,0) {11}}
\put (156,45) {\line(1,1) {14}}
\put (170,59) {\line(3,2) {23}}
\put (198,-4) {\line(3,2) {56}}
\put (259,37) {\line(3,2) {53}}
\put (199,77) {\line(3,2) {54}}
\put (259,117) {\line(3,2) {54}}

\put (198,154) {\line(3,-2) {56}}
\put (259,113) {\line(3,-2) {53}}
\put (199,73) {\line(3,-2) {54}}
\put (259,33) {\line(3,-2) {54}}

\put (319,154) {\line(3,-2) {54}}
\put (319,73) {\line(3,-2) {54}}
\put (317,-4) {\line(3,2) {56}}
\put (319,77) {\line(3,2) {54}}

\thicklines

\multiput(42,167) (60,0){6}{{\large $\wedge$}}
\multiput(42,168) (60,0){6}{{\large $\wedge$}}
\multiput(42,169) (60,0){6}{{\large $\wedge$}}

\multiput (309,12) (0,40){4}{\large $< $}
\multiput (310,12) (0,40){4}{\large $< $}
\multiput (311,12) (0,40){4}{\large $< $}

\thinlines

\put (108,71) {{\Large \it a}}
\put (405,92) {{\Large \it q}}
\put (400,52) {{\Large $ R^{-1} p$}}
\put (405,132) {{\Large \it h}}
\put (405,12) {{\Large \it h}}

\put (42,-35) {{\Large \it v}}
\put (102,-35) {{\Large \it v}}
\put (162,-35) {{\Large \it v}}
\put (222,-35) {{\Large \it v}}
\put (282,-35) {{\Large \it v}}
\put (342,-35) {{\Large \it v}}

\multiput(45,-15)(0,5){38}{.}
\multiput(105,-15)(0,5){38}{.}
\multiput(165,-15)(0,5){38}{.}
\multiput(225,-15)(0,5){38}{.}
\multiput(285,-15)(0,5){38}{.}
\multiput(345,-15)(0,5){38}{.}
 \end{picture}
\vspace{1.5cm}
\caption{\footnotesize Second picture of the function $F_{pq}(a)$.}
\label{Fpq2}
\end{figure}

The probability $F_{pq}(a)$ is unchanged
by this sequence, so we can alternatively use  Figure \ref{Fpq2}
to define $F_{pq}(a)$. We can interchange the rapidity half-lines
$q, R^{-1}p$ therein without changing $F_{pq}(a)$, so
$F_{pq}(a) = F_{p'q'}(a)$, where $q'= R^{-1}p, R^{-1}p' = q$.
Hence
\be \label{fr43}
F_{pq}(a) \eq F_{Rq, R^{-1} p}(a) \comma \ee
which is the relation (15) of \cite{RJB98}.

Alternatively, one can similarly rotate and reverse the left 
half-line anticlockwise round and above $a$ to lie above 
$q$, with $p$ replaced by $Rp$ and again 
with both arrows pointing to the left. This is the configuration of 
Figure \ref{Fpq3}. It too can be used for the definition of 
$F_{pq}(a)$.

Now if we pass the half-line $Rp$ under $q$ we obtain a different 
figure, where $a$ is connected to the other sites only via a single 
edge of $\cal L$, to a site $b$ to the right of $a$. This edge
passes though the intersection of the half lines $Rp, q$ and has
weight $\Wb_{Rp,q}(b-a)$. If we remove it, we regain 
Figure \ref{Fpq3}, but with $q$ interchanged with $Rp$ and $a$
replaced by $b$. Hence 
\be\label{FR3}
F_{pq}(a) \eq  \xi_{pq} \sum_{b=0}^{N-1} \Wb_{Rp,q}(b-a) \,
 F_{R^{-1} q, Rp }(b) \comma \ee
where $\xi_{pq}$ is independent of $a$. With a slight change 
of $\xi_{pq}$, this is the relation (16) of \cite{RJB98}.


\setlength{\unitlength}{1pt}
\begin{figure}[hbt]
\begin{picture}(420,260) (0,0)

\multiput(30,15)(5,0){73}{.}
\multiput(135,55)(5,0){52}{\bf .}
\multiput(135,95)(5,0){52}{\bf .}
\multiput(30,135)(5,0){73}{.}
\put (130,94) {\bf .}
\put (125,90) {\bf .}
\put (123,85) {\bf .}
\put (122,80) {\bf .}
\put (130,56) {\bf .}
\put (125,60) {\bf .}
\put (123,65) {\bf .}
\put (122,70) {\bf .}
\put (120,73) {\line(1,0) {6}}
\put (120,77) {\line(1,0) {6}}
\thicklines

\multiput(42,167) (60,0){6}{{\large $\wedge$}}
\multiput(42,168) (60,0){6}{{\large $\wedge$}}
\multiput(42,169) (60,0){6}{{\large $\wedge$}}

\multiput (309,12) (0,40){4}{\large $< $}
\multiput (310,12) (0,40){4}{\large $< $}
\multiput (311,12) (0,40){4}{\large $< $}

\thinlines


\put (132,61) {{\Large \it a}}
\put (400,52) {{\Large \it q}}
\put (400,92) {{\Large \it Rp}}
\put (400,12) {{\Large \it h}}
\put (400,132) {{\Large \it h}}
\put (143,75) {\circle{8}}
\put (144,80) {\line(3,2) {48}}
\put (198,116) {\line(3,2) {54}}
\put (195,114) {\circle{7}}
\put (255,154) {\circle*{7}}

\put (198,36) {\line(3,2) {54}}
\put (258,76) {\line(3,2) {54}}
\put (318,116) {\line(3,2) {54}}
\put (255,74) {\circle*{7}}
\put (315,114) {\circle{7}}

\put (258,-4) {\line(3,2) {54}}
\put (318,36) {\line(3,2) {54}}
\put (315,34) {\circle{7}}

\put (144,70) {\line(3,-2) {48}}
\put (198,34) {\line(3,-2) {54}}
\put (195,36) {\circle{7}}
\put (255,-4) {\circle*{7}}
\put (375,76) {\circle*{7}}
\put (375,-4) {\circle*{7}}
\put (375,156) {\circle*{7}}

\put (198,114) {\line(3,-2) {54}}
\put (258,74) {\line(3,-2) {54}}
\put (318,34) {\line(3,-2) {54}}

\put (258,154) {\line(3,-2) {54}}
\put (318,114) {\line(3,-2) {54}}

\put (126,162) {\line(3,-2) {67}}
\put (122,165) {\circle*{7}}
\put (75,75) {\circle*{7}}
\put (78,78) {\line(1,2) {42}}
\put (78,72) {\line(1,-2) {42}}
\put (126,-12) {\line(3,2) {67}}
\put (122,-15) {\circle*{7}}

\put (28,165) {\circle*{7}}
\put (72,78) {\line(-1,2) {42}}
\put (72,72) {\line(-1,-2) {42}}
\put (28,-15) {\circle*{7}}

\put (42,-40) {{\Large \it v}}
\put (102,-40) {{\Large \it v}}
\put (162,-40) {{\Large \it v}}
\put (222,-40) {{\Large \it v}}
\put (282,-40) {{\Large \it v}}
\put (342,-40) {{\Large \it v}}

\multiput(45,-25)(0,5){40}{.}
\multiput(105,-25)(0,5){40}{.}
\multiput(165,-25)(0,5){40}{.}
\multiput(225,-25)(0,5){40}{.}
\multiput(285,-25)(0,5){40}{.}
\multiput(345,-25)(0,5){40}{.}
 \end{picture}
\vspace{1.5cm}
\caption{\footnotesize Third picture of the function $F_{pq}(a)$.}
\label{Fpq3}
\end{figure}

\subsubsection*{Domains}

If $x_p, x_q,  y_p, y_q, \omega x_p$ all lie on the unit circle
in an anti-clockwise ordered sequence, then the functions
$W_{pq}(n), \Wb _{pq}(n)$ are real and positive. This is the case 
we have in mind in this paper: it ensures that $\kappa$  and 
the generalized order parameters  
$G_{pq}(r)$ exist in the thermodynamic limit of a large lattice
and are continuous and infinitely differentiable. 

For other values of $p, q$ we define the 
functions to be the analytic continuations from this physical 
regime. To fix our ideas, it is helpful to consider the low-temperature
case, when $k'$ is small. If $\mu_p$ is of order $k'$, then
$x_p$ is free to take most values in the complex plane, other
than those near the $N$ roots of unity 
$1, \omega, \ldots , \omega^{N-1}$, while $y_p$ has to be near 
such a root of unity. If $\mu_p$ is of order ${k'}^{-1}$, then
it is $y_p$ that is free, $x_p$ that is constrained. In the  
Table \ref{domains} we show four such ``domains''. They form 
part of an increasing sequence: if $p$ is in domain 
${\cal D}_r $, then $Rp$ is in domain ${\cal D}_{r+1}$.  

We also introduce the concept of a ``bridge'' ${\cal B}_r$
linking adjacent domains ${\cal D}_r$ and ${\cal D}_{r+1}$.
On such a bridge $\mu_p^N$ is greater than ${\rm O}(k')$
and less than  ${\rm O}({k'}^{-1})$, and $x_p$, $y_p$ are 
constrained  by both ${\cal D}_r$ and ${\cal D}_{r+1}$.
For instance, if $p$ is on the bridge ${\cal B}_1$, then 
$x_p \simeq 1$, $y_p \simeq \omega$.


\begin{table}[hbt]
\begin{center}
\begin{tabular}[t]{| c | c | c | c| }
\multicolumn{3}{c} {} \\
\hline
Domain & $ x_p$ & $y_p$ & $\mu_p^N$ \\ \hline
${\cal D}_{-1} $ &  $\omega^{-1}$ & free &  O(${k'}^{-1}$) \\
${\cal D}_0 $ &  free & 1 &  O($k'$) \\
${\cal D}_1 $   &  1 & free &  O(${k'}^{-1}$)  \\
${\cal D}_{2} $   &  free & $  \omega $ &   O($k'$)  \\
\hline
\end{tabular}
\end{center}
\caption{\footnotesize Some physical rapidity domains}
\vspace{0.5cm}
\label{domains}
\end{table}

If $p$ is in ${\cal D}_r$ and $q$ is on the bridge 
${\cal B}_r$, then  $W_{pq}(n), \Wb _{pq}(n)$
are small for $n \neq 0 $. The same is true if $p$ is on 
${\cal B}_r$ and $q$ is in ${\cal D}_{r+1}$. The 
dominant contribution to the partition 
function is then from all spins being zero. One can develop
the usual series expansions for $\kappa$ for any such case: the
results are all analytic continuations from the
physical regime. We extend our terminology by referring to these 
cases as ``physical''. 

An adjacent ``near-physical'' case is when $p,q \inn 
{\cal D}_r$. Then
$W_{pq}(n) = {\rm O}(1)$, $\Wb_{pq}(n) = {\rm O} ({k'}^{\alpha})$,
where $\alpha = 2 \, \mod (N-r,N)/N$ if $r$ is odd, 
$\alpha = 2 \, \mod (r,N)/N$ if $r$ is even.\footnote{To derive these
formulae, one may have to use the alternative but equivalent form
(\ref{Wbalt}) of (\ref{WWbb})}.
The
$W_{pq}(n), \Wb_{pq}(n)$ are therefore no bigger than 
$W_{pq}(0), \Wb_{pq}(0)$. If this occurs on only one or two 
rapidity lines (or half-lines), while the other (background)
rapidities are as in the previous paragraph, then we still expect
the thermodynamic limit to exist.


The other  ``near-physical'' case is when $p \inn 
{\cal D}_r$, $q \inn
{\cal D}_{r+1}$. This has the same behaviour, except with $W_{pq}(n)$
interchanged with  $\Wb_{pq}(n)$. Again we expect the thermodynamic 
limit to exist if this occurs on only one or two rapidity lines.

We focus attention on the low-temperature case when
$k'$ is small. However, we do not expect any discontinuities
or non-analyticities for $0 < k' < 1$, so expect the above
remarks to generalize to $0 < k' < 1$. For instance in
${\cal D}_1$ we take $|\mu_p| >1$. Then $x_p$ is constrained
to a near-circular region enclosing the point $x_p = 1$
with $|\arg (x_p) | < \pi/(2 N) $, while $y_p$ lies anywhere 
in the complex plane except $N$ such regions surrounding the points
$1, \omega , \ldots , \omega^{N-1}$.\cite[Fig.3]{RJB03}

We do need to consider whether the pictures of Figures \ref{Fpq1},
\ref{Fpq2}, \ref{Fpq3} all correspond to the same analytically
continuous function.
For the physical regime (all Boltzmann weights positive real),
if $p \inn {\cal}_r$, then $q$ should be in domains
${\cal D}_r$, ${\cal D}_{r-1}$, ${\cal D}_{r+1}$, respectively.
Then the thermodynamic limits of the partition functions of 
each figure will exist. They  will be analytic continuations of 
one another, but will lie on different Riemann sheets.

Here we want to keep $p, q$ lying in ${\cal D}_1$ for all three
figures. For Figure \ref{Fpq1}, this is the obvious analytic
continuation of the physical case. We must ask whether we expect
the partition functions of the other two figures to converge,
and if so whether they will analytic continuations of the first.

The answer to both questions is yes. We start with
\be 
p, q, h \inn {\cal D}_1 \sep  v \inn {\cal B}_0 \period \ee
Consider the process outlined 
above for obtaining Figure \ref{Fpq2}. When one rotates and reverses
the half-line $p$ to the lower vertical position, it then
intersects the background horizontal lines with edge weight 
functions $W_{p',q'}, \Wb_{p',q}$, where $p' = R^{-1}p, q' = h$. 
Thus $p' \inn {\cal D}_0$, $q' \inn {\cal D}_1$. This is the second
of the near-physical cases discussed above. The thermodynamic limit
will still exist.

We then shift the background rapidities $v,h$ one half-step 
down to ${\cal D}_0$, ${\cal B}_0$. The background weights remain 
physical, 
as are now those on the half-line  $R^{-1}p$. Now those on q
have weights $W_{vq}, \Wb_{vq}$, where
$v \inn {\cal D}_0$, $q \inn {\cal D}_1$. This is the same 
near-physical case, so the  thermodynamic limit
still exists.

Finally we rotate the line $R^{-1}p$ through one more right-angle
to assume the horizontal position of Figure \ref{Fpq2}. Then the
weights on it become  $W_{v,R^{-1}p}, \Wb_{v,R^{-1}p}$, and
$v, R^{-1} p \inn {\cal D}_0$. This is the first of the near-physical 
cases, so again the  thermodynamic limit exists.

Thus we can go continuously from the $F_{pq}(a)$ of Figure \ref{Fpq1} 
to that of Figure \ref{Fpq2}, keeping the partition function
convergent. We therefore expect both figures to give the same 
analytic function.

A similar argument applies to the rotations and reversals 
necessary to go from Figure \ref{Fpq1} to Figure \ref{Fpq3}.
Now we shift  $v,h$ one half-step  up to ${\cal D}_1$, 
${\cal B}_1$ after reversing $p$ to $Rp$. Again, the
background weights remain physical, while those on the
two half-lines are either physical or near-physical. 
We expect both figures to give the same 
analytic function.


\subsubsection*{The Fourier transform ratio $G_{pq}(r)$}
As in \cite{RJB98}, we define
\be \label{fourF}
\tilde{F}_{pq}(r) \eq \sum_{a=0}^{N-1} \omega^{r a} 
\, F_{pq}(a) \comma \ee
\be \label{defGpq}
G_{pq}(r) \eq \tilde{F}_{pq}(r) /\tilde{F}_{pq}(r-1)  \period \ee
Then the above equations (\ref{FRR}) - (\ref{FR3}) yield
\be \label{firstfr}
G_{Rp,Rq}(r) \eq 1/G_{pq}(N-r+1) \comma \ee
\be G_{RSp,RSq}(r) \eq 1/G_{qp}(N-r+1) \comma \ee
\be G_{pq}(r) \eq  G_{Rq, R^{-1} p}(r) \comma \ee
\be \label{fr4}
G_{pq}(r) \eq  \frac{c_p a_q - a_p c_q \, \omega^r}
{b_p d_q - d_p b_q \, \omega^{r-1}} \; G_{R^{-1}q, R p}(r)  \comma \ee
and {from} (\ref{defGpq}),
\be \label{fr5}
\prod_{r=1}^N G_{pq}(r) \eq 1 \period \ee

Also, 
\be \label{lastfr}
G_{Mp,q}(r) \eq G_{p,M^{-1} q}(r) \eq G_{pq}(r+1) \comma \ee
where $M$ is the rapidity operator such that
\be \label{defM}
\{ a_{Mp}, b_{Mp}, c_{Mp}, d_{Mp} \} \eq 
\{ a_p, \omega^{-1} b_p, \omega^{-1} c_p, d_p \} \period  \ee

A significant point that we shall use is that if $q=p$, then
the break in the rapidity line in Figure \ref{Fpq1}
disappears: the two half-lines become one, and then they {\em can} be
removed to infinity. In the thermodynamic limit the probability 
$F_{pp}(a)$ is therefore independent of $p$, so it is also true that
\be \label{peqq}
G_{pp}(r) \eq {\rm independent \; \; of \; \;}p \period \ee
Since $F_{pp}(a)$ is the probability that the central spin
has value $a$, 
\be \label{calcMr}
{\cal M}_r \eq \langle \omega^{r a} \rangle \eq 
\tilde{F}_{pp}(a)/
\tilde{F}_{pp}(0) \eq G_{pp}(1) \cdots G_{pp}(r) \comma \ee
so a knowledge of $G_{pq}(r)$ is certainly sufficient to determine
the order parameter ${\cal M}_r$.

Equations (\ref{firstfr}) - (\ref{lastfr}) can be regarded as 
functional relations for the functions $G_{pq}(r)$. In \cite{RJB98}
we showed how they can be solved (using an analyticity 
assumption) for the Ising case $N=2$.  We also showed how an obvious 
generalization of this result to $ N >2 $ satisfies the functional 
relations for 
$G_{pq}(r) G_{Rq,Rp}(r)$, but is {\em wrong}. (It disagrees with the 
known series expansions for ${\cal M}_r $.) This is a salutary lesson
that these relations do not by themselves determine 
$G_{pq}(r)$: one has to input the correct analyticity properties.




\section{Low-temperature limit}

Now we briefly consider the low-temperature
limit, when $k'$ is small, using the picture of Figure \ref{Fpq1}. 

As in \cite{RJB98,RJB98b}, we focus on  the case when
$p,q \inn {\cal D}_1$, when
\be \label{regimeA1}
x_p , x_q \simeq 1 \sep  \mu_p, \mu_q = {\rm O} ({k'}^{-1/N})  \ee
so $y_p, y_q$ are arbitrary, of order unity.
We also take $v \inn {\cal B}_0$, $h \inn {\cal D}_1$.
This includes the physical regime where all the 
Boltzmann weights are real and positive.

The dominant contribution to $Z(a)$ then
comes from the configuration where all spins other than $a$
are zero. Then
\be
F_{pq}(a) \eq W_{vp}(N-a) \, \Wb_{v'q}(N-a) \, \Wb_{vh}(a) \,
 W_{v'h}(a)  \comma \ee
allowing the two vertical rapidities to have different values $v, v'$.
Since $\Wb_{pq}(N) = 1$, we can divide the relation 
(\ref{WWbb}) by $\Wb_{pq}(N)$ to obtain
\be  \label{Wbalt}
\Wb_{pq}(n) \eq  (\mu_p \mu_q)^{n-N}  \prod_{j=n+1}^N 
\frac {y_q - \omega^j y_p} {\omega x_p - \omega^j x_q}  \period \ee
Using this, (\ref{WWba}) and (\ref{xymu}) (with $k = 1$), we find 
that $F_{pq}(0) =1$, while
for $ 1 \leq a \leq N-1$,
\be \label{FpqzA1}
F_{pq} (a) \eq \frac{{k'}^2 (\mu_p/\mu_q)^a }{N^2 
(1-\omega^a)(1-\omega^{-a}) }\; \prod_{j=1}^a \frac {1-\omega^{j-1} t_q}
 {1-\omega^{j-1} t_p} \period \ee
This satisfies the relation (\ref{FRS}).

As expected, the background rapidities $v, v', h$ cancel out of these
expressions. Also, when $q=p$,  $F_{pq} (a) $ is independent of $p$.
{From} (\ref{fourF}),
\be \langle \omega^{r a} \rangle \eq \frac{\tilde{F}_{pp}(r)}
{\tilde{F}_{pp}(0)} \eq 1 - {k'}^2 r (N-r)/(2 N^2 ) + 
{\rm O }({k'}^4) \comma \ee
in agreement with (\ref{conj}) and (\ref{conj2}). (The second-order 
terms in $F_{pq}(0)$ cancel out of this calculation.)

For $N=3$ the author  has obtained series expansions for $F_{pq}(a)$
to order ${k'}^8$.\cite[eq. 48]{RJB98b}.




\section{A solvable special case}

We have not determined $G_{pq}(r)$ for general values 
of the two half-line rapidities $p$ and $q$. What we have done is
obtain it for the case when $\{ a_q, b_q, c_q, d_q \} 
=  \{ a_p, \omega b_p, c_p, d_p \}$, i.e.
\be \label{spcase}
x_q = x_p \sep y_q = \omega y_p \sep \mu_q = \mu_p \period \ee
The calculation and the result are very similar to the calculation
of the free energy of the $\tau_2(t_q)$ model.

We have given the expressions (\ref{WWb}) for the Boltzmann
weights in terms of the original rapidity variables 
$a_p, b_p, c_p, d_p$ because they make it clear that
the weights remain finite and non-zero when $b_p, b_q$ become zero. 
This corresponds to taking $y_p = y_q = 0$. Hence the special
case (\ref{spcase}) then intersects the desired case $q=p$. 
{From} (\ref{peqq}), this is sufficient to determine $G_{pp}(r)$
and hence the order parameter ${\cal M}_r$.

In this section we shall consider the three pictures 
in Figures \ref{Fpq1}, 
\ref{Fpq2}, \ref{Fpq3} of the function $F_{pq}(a)$. They all 
give the same function $F_{pq}(a)$ in the limit of an infinite 
lattice. However, they do differ for a finite lattice. We shall 
use this difference to manifest different properties of 
$F_{pq}(a)$.

 We shall then put these properties
together to make what we believe to be a plausible and 
correct assumption as to the analyticity properties of 
the Fourier transform ratio  $G_{pq}(r)$ for $ 0 <k' <1 $.
This assumption is the key to derivation of this paper.

One function that we shall use is
\be \label{epsr}
 \epsilon (r) \eq 1 - N \delta_{r,0} \comma \ee
where $\delta_{r,0}$ is to be interpreted modulo $N$. Thus
$ \epsilon (0) =  \epsilon (N) =1-N$, and 
$\epsilon (r) = 1 $ for $r = 1, \ldots , N-1$.

\subsection*{First picture}

Consider the definition (\ref{FZZ}) of $F_{pq}(a)$, where
$Z(a)$ is the partition function of the lattice shown in Figure
\ref{Fpq1}. For the moment ignore the
restrictions (\ref{spcase}) and allow $p$, $q$ to be general
rapidities. For
a finite lattice (with boundary spins fixed to zero),
it is readily seen that
\be \label{fv1}
F_{pq}(a) \eq (\mu_p/\mu_q)^a \times \; \; ({\rm  rational 
\; \; function \; \; of \; \;} x_p, y_p, x_q, y_q )
\period \ee

Consider $\mu_p, x_p, y_p$ as functions of the complex variable
$t_p$. {From} (\ref{xymu}) they are multi-valued functions,
with $N$ branch cuts ${\cal C}_0, \ldots , {\cal C}_{N-1}$.
The cut  ${\cal C}_i$ is along the straight-line segment
$(\omega^i \rho , \omega^i/\rho )$, 
where $\rho = [(1-k')/(1+k')]^{1/N}$. For $N=3$, 
these cuts are shown in Figure
\ref{brcuts}. On them $|\mu_p| = 1$. Here we concentrate our
attention on the case where $p \inn {\cal D}_1$, which is when
\be \label{domD1}
|\mu_p | > 1 \sep |\arg{x_p} | < \pi/(2 N) \period \ee
For $k'$ small, $x_p$ is restricted to a small region $\cal R$
round $x_p=1$. We shall say $x_p \in {\cal R}$, or 
simply $x_p \simeq 1 $. On the other hand, $y_p$ can lie
almost anywhere in the complex plane, being excluded only 
from $\cal R$ and corresponding small regions round 
$y_p = \omega, \ldots, \omega^{N-1}$.

For a finite lattice, it follows from (\ref{fv1}) that
$F_{pq}(a) $ is also a multi-valued function of $t_p$. To make it
single-valued, we must restrict $t_p$ to the cut plane
Figure \ref{brcuts}.

Now consider the low-temperature limit of section 5.
Making the substitutions (\ref{spcase}) into (\ref{FpqzA1}),
we observe that there are significant cancellations. We obtain
\be \label{FpqzA2}
F_{pq} (a) \eq \frac{{k'}^2  }{N^2 
(1-\omega^a)(1-\omega^{-a}) } \; \frac {1 -\omega^{a} t_p}
 {1- t_p}   + {\rm O} ({k'}^4) \period \ee
{From} (\ref{fourF}), (\ref{defGpq}) it follows that
\bd G_{pq}(r)  \eq 1 - \frac{ (N-1-2r) {k'}^2}{2 N^2}  - 
\frac{\er \, {k'}^2}{N^2 (1-t_p)} + {\rm O} ({k'}^4) \ed
for  $ r = 0 ,\ldots , N-1 $.

To this order there is no evidence of any zero or singularity
in $G_{pq}(r)$ for any value of $t_p$ other than one. This
singularity is consistent with there being a branch cut 
on the real axis, as in Figure \ref{brcuts}, but at this very 
preliminary stage there is no evidence that the other
$N-1$ branch cuts, or any other singularities,  occur.




\setlength{\unitlength}{1pt}
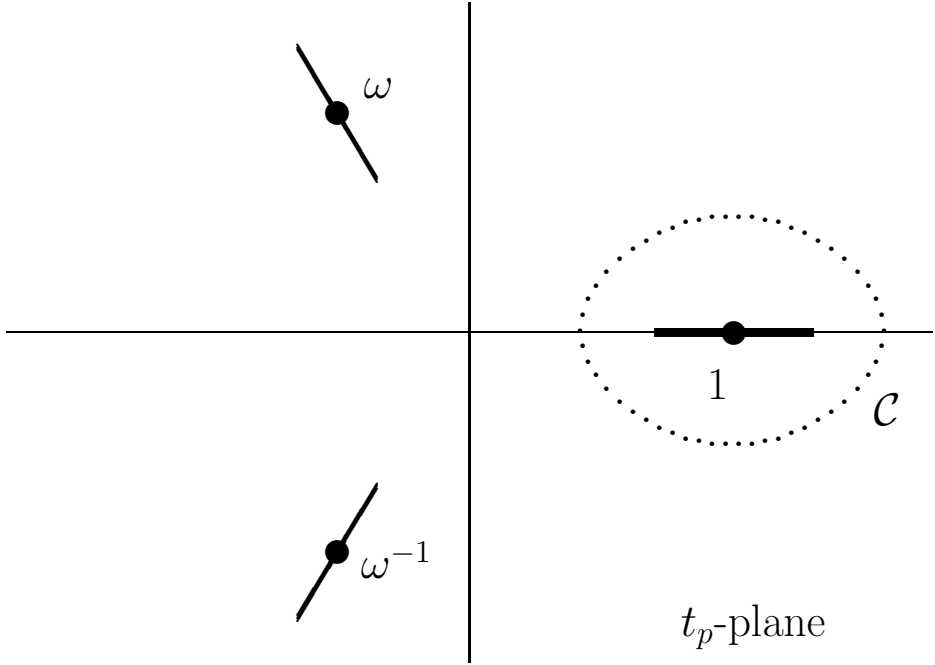
\begin{figure}[hbt]
\begin{picture}(420,260) (0,0)

\put (50,125) {\line(1,0) {350}}
\put (225,0) {\line(0,1) {250}}

\put (325,125)  {\circle*{9}}
\put (175,208)  {\circle*{9}}
\put (175,42)  {\circle*{9}}

\put (315,100)  {\Large 1}
\put (185,214)  {\Large $\omega$}
\put (184,32)  {\Large $\omega^{-1}$}

\put (378,90) {{\Large {$\cal C$}}}

\put (305,10) {{\Large {$t_p$-plane}}}
\thicklines
\put (295,124) {\line(1,0) {60}}
\put (295,125) {\line(1,0) {60}}
\put (295,126) {\line(1,0) {60}}

\put (265,125) {\bf .}
\put (266,131) {\bf .}
\put (268,137) {\bf .}
\put (271,142) {\bf .}
\put (275,147) {\bf .}
\put (280,152) {\bf .}
\put (285,156) {\bf .}
\put (290,159) {\bf .}
\put (295,162) {\bf .}
\put (300,164) {\bf .}
\put (305,166) {\bf .}
\put (310,167) {\bf .}
\put (315,168) {\bf .}
\put (320,168) {\bf .}
\put (325,168) {\bf .}
\put (330,168) {\bf .}
\put (335,167) {\bf .}
\put (340,166) {\bf .}
\put (345,164) {\bf .}
\put (350,162) {\bf .}
\put (355,159) {\bf .}
\put (360,156) {\bf .}
\put (365,152) {\bf .}
\put (370,147) {\bf .}
\put (374,142) {\bf .}
\put (377,137) {\bf .}
\put (379,131) {\bf .}
\put (380,125) {\bf .}

\put (266,119) {\bf .}
\put (268,113) {\bf .}
\put (271,108) {\bf .}
\put (275,103) {\bf .}
\put (280,98) {\bf .}
\put (285,94) {\bf .}
\put (290,91) {\bf .}
\put (295,88) {\bf .}
\put (300,86) {\bf .}
\put (305,84) {\bf .}
\put (310,83) {\bf .}
\put (315,82) {\bf .}
\put (320,82) {\bf .}
\put (325,82) {\bf .}
\put (330,82) {\bf .}
\put (335,83) {\bf .}
\put (340,84) {\bf .}
\put (345,86) {\bf .}
\put (350,88) {\bf .}
\put (355,91) {\bf .}
\put (360,94) {\bf .}
\put (365,98) {\bf .}
\put (370,103) {\bf .}
\put (374,108) {\bf .}
\put (377,113) {\bf .}
\put (379,119) {\bf .}

\put (160,16) {\line(3,5) {30}}
\put (160,17) {\line(3,5) {30}}
\put (160,18) {\line(3,5) {30}}

\put (160,234) {\line(3,-5) {30}}
\put (160,233) {\line(3,-5) {30}}
\put (160,232) {\line(3,-5) {30}}

\thinlines

\ \end{picture}
\vspace{1.5cm}
\caption{\footnotesize The cut $t_p$-plane for $N=3$, showing the contour 
$\cal C$.}
\label{brcuts}
\end{figure}

\subsection*{Second picture}

Now we consider the picture of  Figure \ref{Fpq2} for $Z(a)$
and write  
\be p' = R^{-1} p   \ee
for the rapidity of the lower half-row. The rapidity of the upper 
half-row is $q$.

Then 
$ x_p = y_{p'} ,  y_p = \omega x_{p'}, \mu_p = 1/\mu_{p'}$, so
from (\ref{spcase}) the rapidities of the two rows are related by
\be \label{2ndcondns}
x_q = y_{p'} \sep y_q = \omega^2 x_{p'} \sep \mu_q = 1/\mu_{p'} 
\period \ee

For the moment, take the lattice to be finite, of half-width $s$.
In Figure \ref{Fpq2} we have shown the sites of ${\cal L}$
immediately above and below 
the two  half-rapidity lines (including the left-hand site with spin 
$a$) as open circles. All other sites are shown as filled circles. 

If we duplicate the spin $a$ as two open circles at 
the same heights as the others, these open circle spins
correspond precisely to the spins in Figure \ref{tworows}. 
The spins between them lie at the centre of
stars, as in Figure \ref{Uabcd}.
Summing over them, we obtain precisely the weight (\ref{combwt}),
 with $\sigma_1, \sigma'_1 = a$. Because the boundary 
spins are fixed to zero, we also have $\sigma_s, \sigma'_s = 0$.

Comparing (\ref{2ndcondns}) with (\ref{rowreln}), we see that 
we have 
\be k = 0  \sep \ell = 2 \sep  j = 2  \comma \ee
so we can use the results of the $j=2$ sub-section of Section 3
for the contribution to each term in $Z(a)$ of edges crossing 
the half-lines 
$R^{-1}p, q$.  Thus $Z(a)$ is a sum of expressions (\ref{combwt})
(with coefficients independent of $p, q$), so from (\ref{I1})
{\em $Z(a)$ is a polynomial in $t_p$ of degree $s-1$}. It has no 
other dependence on $p$.

Again we consider the low-temperature case (\ref{regimeA1}).
The dominant contribution to any $Z(a)$ comes from 
the configuration where all the remaining spins other than 
$a$ are zero.
Also, the other four edges incident to spin $a$ give a weight
that is unity when $a=0$, while when $a \neq 0$  to leading order
their contribution is ${k'}^2/[N^2 (1-\omega^a)(1-\omega^{-a})]$.

Using Table \ref{Uwts}, with $p$ replaced by 
$p'$ and $y_v = y_{v'}=1$, 
and noting from the equations of this sub-section that 
$\omega t_{p'} = t_p$, it follows that
\bd
Z(0) \eq  (1-t_p)^{s-1} + {\rm O}({k'}^2) \comma \ed
and for $a \neq 0 $
\be Z(a) \eq \frac{{k'}^2 (1-\omega^a t_p)(1-t_p)^{s-2}}
{N^2 (1-\omega^a)(1-\omega^{-a})}
+ {\rm O}({k'}^4) \period \ee
Again we obtain the leading-order result (\ref{FpqzA2})
for $F_{pq}(a)$, but now we see also
that, for $s$ finite and $a \neq 0 $,
\be
\tilde{F}_{pq}(a) \eq \tilde{Z}(a)/\tilde{Z}(0) \comma \ee
where the $\tilde{Z}(a)$ are polynomials in $t_p$ of degree 
$s-1$, equal to $(1-t_p)^{s-1}$ when $k' = 0$.

Hence, by continuity, for sufficiently small $k'$ the zeros
of the polynomials  $\tilde{Z}(a)$ must lie close to $t_p = 1$.

As we take the limit of $s$ large, we only expect singularities
to occur in the vicinity of these zeros.
This reinforces the suggestion of the last sub-section that
in the limit of a large lattice,
$\tilde{F}_{pq}(a)$ and $G_{pq}(r)$  may be analytic functions of $t_p$
except for singularities in some region surrounding $t_p = 1$.
Again, there is no evidence of the branch cuts in Figure \ref{brcuts},
other than the cut on the real positive axis.



\subsection*{Third picture}

Finally we consider the picture of  
Figure \ref{Fpq3} for $Z(a)$, and
write  
\be p' = q \sep q' = R p   \ee
for the rapidities of the half-rows immediately below and above 
 the spin $a$. 

Then 
$ x_{q'} = y_{p} ,  y_{q'} = \omega x_{p}, \mu_{q'} = 1/\mu_{p}$, so
from (\ref{spcase}) the rapidities of the two rows are related by
\be \label{3rdcondns}
x_{q'} = \omega^{-1} y_{p'} \sep y_q = \omega x_{p'} \sep 
\mu_q = 1/\mu_{p'}  \period \ee

Again we initially take the lattice to be finite, of half-width $s$.
The  open circle spins  in Figure \ref{Fpq3}, other than $a$,
correspond  to the spins in Figure \ref{tworows}. Let the lower such 
spins be $\sigma_1, \ldots, \sigma_s$, the upper ones
$\sigma'_1, \ldots, \sigma'_s$.
Summing over the spins between them, we obtain the weight 
(\ref{combwt}), with the boundary spins  $\sigma_s, \sigma'_s = 0$.
We also obtain the weight of the two edges incident to $a$. {From} 
(\ref{fourF}) and (\ref{defVvpq}), these contribute a further
weight $V_{vq'p'}(-\sigma'_1, -\sigma_1;-r)$ to $\tilde{F}_{pq}(r)$. 
Using (\ref{defJr}), we see that the total factor contributed
by the edges that cross  the lines $q, Rp$ is 
$J_r(\sigma_1, \ldots \sigma'_s)$.

{From} (\ref{3rdcondns}), 
\be k=-1 \sep \ell = N+1 \sep j = N \comma \ee
so we can use the results of the last sub-section of section 3, with
$\nu = N-1$. Summing over all the spins other than $a$ in Figure
\ref{Fpq3}, we see from (\ref{J1}) that
\be \label{tF3}
x_p^{m(r)}  \, \tilde{F}_{pq}(r) \eq {\cal A}_r/{\cal A}_0  \comma \ee
where
\be m(r) \eq \mod (N-r-1,N)  \ee 
and each ${\cal A}_r$ is a weighted sum of functions
${\cal J}_r$, with coefficients independent of $p$.
Thus ${\cal A}_r$ is a polynomial in $t_p$ of degree 
$(N-1) (s-1)$.
This is the complete dependence on $p$.

When $k' = 0$, then $x_v = x_{v'} = 1$ and the  only contribution 
to the sum-over-states is
from all spins being zero, so from (\ref{Jrlt})
\be {\cal A}_r \eq \prod_{i=1}^{N-1} (1-\omega^i t_p)^{s-1} \period \ee
{From} continuity, for sufficiently small $k'$ 
we see that all the zeros of the polynomials ${\cal A}_r$ must be 
clustered round the $N$th roots of unity in the complex $t_p$-plane, 
{\em excluding} $t_p = 1$.

So now we have information that is reciprocal to that of the previous
sub-section. In the thermodynamic limit of $s$ large we 
expect $x_p^{m(r)}  \tilde{F}_{pq}(r)$ to have
singularities near the  $t_p = \omega, \omega^2, 
\ldots , \omega^{N-1}$, but {\em not} near $t_p = 1$.
{From} (\ref{tF3}) and (\ref{defGpq}),
\be \label{G1}
G_{pq}(r) \eq x_p^{\er} {\cal A}_r/ {\cal A}_{r-1} \comma \ee
for $r = 0, \ldots, N-1$.

\subsubsection*{The relation (\ref{fr4}).}
So far we have not used any of the six functional relations
(\ref{firstfr}) - (\ref{lastfr}). Half of them are not
helpful for the present argument, since if the arguments
$p,q$ of the function $G$ on one side of the relation
satisfy our restriction (\ref{spcase}), then those on the other side
do not.

One relation that is helpful is (\ref{fr4}). Set
\be p_1 = R^{-1} q \sep q_1 = Rp \comma \ee
then, using (\ref{Rop}) and (\ref{spcase}), we obtain
\be x_{q_1} = x_{p_1} \sep y_{q_1} = \omega y_{p_1} \sep \mu_{q_1} = 
\mu_{p_1} \comma \ee
so $p_1$ and $q_1$ also satisfy the restriction (\ref{spcase}).
The relation  (\ref{fr4}) becomes
\be G_{pq}(r) \eq \frac{x_p (1-\omega^r)}{y_p  (1-\omega^r)} \; 
G_{p_1,q_1}(r) \comma \ee
which implies
\be \label{nobrcut1}
x_p^{-1}  G_{pq}(r) \eq y_p^{-1}  G_{p_1,q_1}(r) \ee
for $r = 1, \ldots , N-1$. Also, from (\ref{fr5}) it follows that
\be \label{nobrcut2}
x_p^{-\er}  G_{pq}(r) \eq y_p^{-\er}  G_{p_1,q_1}(r)  \ee
for $r = 0, \ldots , N-1$.

If we write $p = \{x_p, y_p, \mu_p \}$, then
$p_1 = \{ y_p, x_p, 1/\mu_p \}$. Hence $p_1$ is obtained from
$p$ by interchanging $x_p$ with $y_p$ and inverting $\mu_p$, 
which leaves $t_p$ interchanged. This is what one obtains by
taking $t_p$ across the branch cut on the positive real axis in
Figure \ref{brcuts} and then returning it to its original value.
The relations (\ref{nobrcut1}), (\ref{nobrcut2})  therefore say that 
the functions on the lhs of the equations also return to their 
original values, i.e. that they do {\em not} have this branch cut in 
the complex $t_p$ plane. This agrees with (\ref{G1}) 
and our observations that we do not expect the ratios of the 
${\cal A}_r$ to have any singularities in the vicinity of $t_p=1$.



\section{Analyticity assumptions}

Define, for $r = 1, \ldots , N$,
\be \label{defH}
H_{pq} (r) = x_p^{-\er}  G_{pq}(r) \period  \ee

The functions $G_{pq}(r)$ are defined by (\ref{fourF}) and
(\ref{defGpq}) in terms of $F_{pq}(a)$. This in turn is proportional to
$Z(a)$, the partition function of the lattice with the broken 
rapidity line. Three equivalent forms of this lattice are shown
in Figures \ref{Fpq1}, \ref{Fpq2}, \ref{Fpq3}.


In the previous section we took $q$ to be related to $p$ by
(\ref{spcase}). We presented evidence
based on Figures \ref{Fpq1} and \ref{Fpq2} to suggest
that for sufficiently small $k'$ the functions $G_{pq}(r)$ are 
analytic and non-zero
in the complex $t_p$-plane, excepting some region surrounding and not 
far from the point $t_p=1$. Then  we used Figure \ref{Fpq3}
to obtain evidence suggesting that the functions 
$H_{pq}(r)$ are analytic and non-zero except near
$t_p = \omega, \ldots , \omega^{N-1}$. We showed that this is 
consistent with the functional relation (\ref{fr4}).

We remarked at the beginning of section 5 that the chiral Potts 
Boltzmann weight functions $W_{pq}(n), \Wb_{pq}(n)$ remain 
finite when $b_q = 0 $ and  $a_q, c_q, d_q \neq 0 $. For 
this reason we neither see nor expect
any singularities in the $G_{pq}(r)$ at $t_p = 0$. The same
is true when $c_q = 0$, implying that the $G_{pq}(r)$ remain
finite and analytic when $t_p \rightarrow \infty$.

We therefore make the following analyticity assumptions, taking 
$p, q$ to be related by (\ref{spcase}), and
regarding $G_{pq}(r), H_{pq}(r)$ as functions of the complex 
variable $t_p$. We take both $p$ and $q$ to lie in the domain
${\cal D}_1$, i.e. to satisfy (\ref{domD1}).
Although our evidence is obtained when $k'$ is 
sufficiently small, we do not expect any non-analyticities in 
$k'$ throughout the  ferromagnetic  regime $0 < k' < 1$.
We therefore expect the assumptions to be true for $0 < k' < 1$.

\vspace{5mm}

ASSUMPTIONS

There exists a closed contour $\cal C$ surrounding the branch cut
on the positive real axis in Figure \ref{brcuts}, such that

i)  the functions $G_{pq}(r)$ are analytic, bounded and non-zero
 outside and on  $\cal C$.

ii)  the functions $H_{pq}(r)$ are analytic and non-zero inside
and on $\cal C$.

\vspace{5mm}

The functional relation (\ref{fr4}) implies that  $H_{pq}(r)$
has no branch cut on the positive real axis, so (ii) follows if
we strengthen (i) to apply to the whole $t_p$-plane except for the 
cut on the positive real axis.

These assumptions are very similar to those we made to calculate
the free energy of the $\tau_2(t_q)$ model by the ``inversion 
relation'' method \cite[p. 419]{RJB03}. In fact we shall find that 
$G_{pq}(r)$ has a similar form to that free energy.

These analyticity properties can be expressed as 
symmetries.\cite{RJB03,RJB03b} For instance,
if we exhibit the dependence of $G_{pq}(r)$ on $x_p, y_p$ by writing 
it as $G_r(x_p,y_p)$, then Assumption (i) implies
\bd
G_r(x_p,y_p) \eq G_r(\omega^j y_p,\omega^{-j} x_p) \comma \ed
for $j = 1, \ldots , N-1$, $(x_p, y_p) \inn {\cal D}_1$ and 
$(\omega^j y_p,\omega^{-j} x_p)$ in a domain adjacent to ${\cal D}_1$.



\section{Calculation of  $G_{pq}(r)$}

The above assumptions are sufficient to calculate  each of
$G_{pq}(0), \ldots , G_{pq}(N-1)$ to within some constant factor.
These factors can then be determined from the property
(\ref{peqq}).

The we can write (\ref{defH}) as
\be \label{xfac}
x_p^{\er}  = G_{pq} (r)/H_{pq}(r) \period  \ee
Then it follows immediately from the above Assumption that
we can obtain the functions $G_{pq} (r), H_{pq}(r)$ by  a Wiener-Hopf
factorization of $x_p^{\er}$. (See for example
eqns. (50) - (53) of \cite{RJB03}.) 

The variables $x_p, \mu_p$ are function of $t_p$: let us write them as 
$x(t_p)$, $\mu (t_p)$.
The function $\log x(t_p)$ is 
single-valued and analytic on $\cal C$ in the $t_p$-plane.
Define
\be \label{defB-}
B_{-}(t) \eq \frac{1}{2 \pi \i } \, 
\oint_{\cal C}  \frac{\log x(s)}{s-t} \, {\rm d}s \comma \ee
where the integration is now round the corresponding curve $\cal C$
in the complex $s$-plane, and $t$ lies outside $\cal C$.
Define $B_{+}(t)$ by the same equation, but taking
$t$ to be inside $\cal C$. In the first case shrink  $\cal C$
to a curve $\cal C_{-}$ just inside $\cal C$, in the second 
expand it to $\cal C_{+}$ lying just outside $\cal C$. Now take $t$ to 
lie between $\cal C_{+}$ and $\cal C_{-}$. Then
the combined curve of integration for $B_{+}(t) - B_{-}(t)$ 
can be deformed to a closed curve surrounding $t$ and close to $\cal C$, 
so by Cauchy' integral formula
\be B_{+}(t) - B_{-}(t) \eq \log x(t) \period \ee

Hence from (\ref{xfac}),  
\be \er B_{-}(t) + \log G_{pq}(r)   \eq   \er B_{+}(t) + \log H_{pq}(r) 
\period \ee

{From} (\ref{defB-})  and our assumption, the lhs is analytic 
and bounded outside $\cal C$. Similarly, the rhs is analytic 
and bounded inside $\cal C$. Each side is therefore entire and bounded,
so be Liouville's theorem each is a constant. Hence
\be  \label{Gpqres1a} 
 \log G_{pq}(r) \eq \; \; {\rm constant} \; \; - \er B_{-}(t_p) 
\comma \ee
where the constant is independent of $p$, but may depend on $r$.


This result is of course consistent  with our assumption, and with the
stronger assumption that  $G_{pq}(r)$  is  analytic and  non-zero in 
the $t_p$-plane, except for the cut along the real axis in Figure 
\ref{brcuts}. The contour $\cal C$ can be shrunk to just surround this 
cut.

Integrating by parts, we can write  (\ref{defB-})
as
\be  B_{-}(t_p ) \eq   \eq - \frac{1}{2 \pi \i } \, 
\oint_{\cal C}  \log (s-t_p) \, \frac{x'(s)}{x(s) } \, {\rm d}s 
\comma \ee

The quantity $\mu (s)^N$ moves once around the unit circle  as $s$ 
moves around $\cal C$ (in the  positive directions). Setting
$\mu (s) = \exp (-\i \theta/N)$ and using 
(\ref{Gpqres1a}), we obtain for 
$r = 1, \ldots , N$,
\be \label{Gfinres}
G_{pq}(r) \eq C_r \,{ S(t_p) } ^{\er }  \comma \ee
where $C_r$ is a constant, 
\be \label{defS}
  \log S(t_p)  \eq - \frac{2}{N^2} \log k   + \frac {1}{2 N  \pi  } \, 
\int_0^{2 \pi}   \frac{k' \e^{\i\theta}}{1-k' \e^{\i\theta}} \, 
\log [\Delta(\theta) - t_p] \,  {\rm d}\theta  \comma \ee
and
\be \Delta( \theta ) \eq [(1-2k' \cos \theta + {k'}^2 )/k^2]^{1/N} 
\period \ee
The $\log k$ term has been included in (\ref{defS}) to ensure
that $S(0) = 1$, choosing the logarithms to ensure that
$S(t_p)$ is positive real when $t_p$ is 
real and off the branch cut. The condition (\ref{fr5}) 
is satisfied provided that 
\be \label{prodC}
C_1 C_2 \cdots C_N \eq 1 \period \ee

\subsubsection*{Calculation of $C_1, \ldots , C_{N}$.}

We still need to determine the constants $C_1, \ldots , C_{N}$.
We noted at the beginning of section 5 that when $y_q = y_p = 0$,
with the restriction (\ref{spcase}), $G_{pq}(r) = G_{pp}(r)$.
In this case $t_p = 0$ and we can evaluate the integral in 
(\ref{Gfinres}), giving $S(0) = 1$ and 
\be \label{rel1}
G_{pp}(r) \eq  C_r \period \ee

Now consider the case when $a_q, \ldots , d_p$ satisfy 
the restrictions given immediately before (\ref{spcase}),
and $c_q = c_p = 0$, so $y_q = y_p = \infty$. It is not true
that $q = p$, but from (\ref{defM}) it is true that
$q = M^{-1} p$.
Now $t_q = t_p = \infty$ and the integral in (\ref{defS})
is zero.
Using (\ref{lastfr}), we therefore have
\be  \label{rel2}
G_{pp}(r+1) = G_{p,M^{-1}p}(r) = C_r \, k^{-2 \er/N^2} \period \ee

Because of (\ref{peqq}), the $G_{pp}(r) $ in these 
last two equations are the same. Eliminating them, we obtain
\be C_{r-1} = k^{2 /N^2}\, C_r  \ee
for $r = 2, \ldots , N$. {From} these equations and
(\ref{prodC}), it follows that
\be \label{Cr}
C_r \eq k^{(N+1-2r)/N^2} \ee
for $r = 1, \ldots, N $. 

This completes the calculation of $G_{pq}(r)$ for 
the case when $p, q$ are related by (\ref{spcase}): it is given 
by (\ref{Gfinres})  and   (\ref{Cr}). 

We also immediately 
obtain from (\ref{rel1}) that
\be \label{Gppres}
G_{pp}(r)  \eq k^{(N+1-2r)/N^2} \ee
for $r = 1, \ldots, N $. Hence from  (\ref{calcMr}),
\be {\cal M}_r \eq \langle \omega^{r a} \rangle \eq k^{r(N-r)/N^2} 
\comma \ee
for $r = 0, \ldots, N$. This verifies the conjecture (\ref{conj}).

\subsubsection*{Relation to  $\tau_2(t_p)$.}
The free energy of the $\tau_2(t_p)$ model is given in
eqn. (39) of \cite{RJB91} and (73) of \cite{RJB03}. These 
expressions are very 
similar to the integral in (\ref{Gfinres}). In fact if we use
eqn. (73) of \cite{RJB03} for  ${\tau}_2(t_p)$, and manifest its 
dependence on $\lambda_q$ by writing it as 
${\tau}_2(t_p, \lambda_q)$, then
\be 
k^{1/N} \, {S(\omega t_p)}^{2N}  \eq \tau_2(t_p,0)/\tau_2(t_p,k')
 \period \ee

\subsubsection*{The product $G(t_p,r) \cdots G(\omega^{N-1}t_p,r)$.}

Exhibit the dependence of  $G_{pq}(r)$ on $t_p$ by writing it as 
$G_r(t_p)$. Consider the product
\be L_r(t_p) \eq G_r(t_p)  G_r(\omega t_p) \cdots 
G_r(\omega^{N-1}t_p) \period \ee
This can be evaluated from (\ref{Gfinres}), giving:
\be \label{Lpr}
 L_r(t_p) \eq D_r  x_p^{\er} \comma \ee
where 
\be \label{Dr}
D_r = C_r^N k^{-\er/N}  = k^{1+\delta_{r,N}-2r/N} \ee
for $r = 1, \ldots , N$.

In fact we could have obtained this result very directly from the 
assumptions of section 6. These imply that  the functions 
$x_p^{-\er}G_r(t_p)$, $G_r(\omega t_p)$, ... ,$G_r(\omega^{N-1}t_p)$
are all analytic on and near the positive real axis. Hence so is
the product $L_r(t_p)/x_p^{\er}$. Since this is unchanged by 
multiplying $t_p$ by $\omega$, none of the branch cuts in Figure 
\ref{brcuts} appear, so it is analytic everywhere. It is 
bounded, so by Liouville's theorem it is  a constant. This verifies 
(\ref{Lpr}), but  does not determine the constant. To do this, repeat 
the argument of (\ref{rel1}) - (\ref{Cr}). {From} (\ref{xymu}),
 $x_p = k^{1/N}$ when
$y_p=0$, and $x_p = k^{-1/N}$ when $y_p = \infty$, so we obtain
\be
G_{pp}(r)^N \eq D_r k^{\er/N} \eq D_{r-1} k^{-\epsilon(r-1)/N} 
\period \ee
{From} (\ref{fr5}), $D_1 D_2 \cdots D_{N} = 1$. It follows that
$D_r$ is indeed given by (\ref{Dr}). It also follows that
$G_{pp}(r)$ is given by (\ref{Gppres}), which verifies the conjecture
(\ref{conj}). This route avoids using the Wiener-Hopf factorization 
and the integral (\ref{Gfinres}).



\section{Other special cases.}

It is natural to ask if we can evaluate $G_{pq}(r)$ when
$p, q$ are related, not by (\ref{spcase}), but by the more
general condition
\be \label{genrestr}
x_q = x_p \sep y_q = \omega^i y_p \sep \mu_q = \mu_p \comma \ee
where $i = 0, \ldots , N-1$.

For such a case, let us introduce an extra index $i$ and write
\be
G_{pq}(r) \eq G_{i \, r}(t_p) \sep  L_{i \,  r}(t_p) \eq  
\prod_{j=0}^{N-1} G_{i \, r}(\omega^j t_p) \period \ee

As a first step, one can use the series (48) of \cite{RJB98b}
to expand $L_{i \, r} (t_p)$ to order ${k'}^8$ for $N=3$. 
Indeed, this is how the author discovered that the case 
(\ref{spcase})
might be tractable. For $i=2$ and $r=0, 2$ we have not observed
anything particularly simple, but for $i=2, r=1$ the series are 
consistent with the conjectures
\be L_{21}(t_p) \eq x_p^2 \sep 
G_{21}(t_p) \eq k^{2/9} S(t_p) S(\omega t_p) \comma \ee
$S(t_p)$ being the function defined by (\ref{defS}).

We can use the general result (\ref{FpqzA1}) to expand
$G_{i \, r}(t_p)$ to first order in ${k'}^2$. For 
$i=0, \ldots , N-1$ and $r = 1, \ldots , N-i$
the results simplify, giving
\bd 
G_{i \, r}(t_p) \eq 1 + \frac{{k'}^2}{ 2N^2} \left( 2r+2i-N-1 -
2 \sum_{j=1}^i
\frac{1}{1- \omega^{j-1} t_p}\right) \comma \ed
which  is  consistent with the formula
\be \label{myconj1}
G_{i \, r} (t_p) \eq k^{(N+1-2r)/N^2} \, S(t_p) S(\omega t_p) \cdots
 S(\omega^{i-1}  t_p) \period \ee
This formula also agrees with (\ref{Gppres}) when $t_p = 0,q=p$, 
and when $t_p = \infty, q = M^{-i}p$. From preliminary calculations we 
suspect (\ref{myconj1}) can be justified 
in the same way that we justified Assumptions (i) and (ii), by arguing 
that $G_{pq}(r)$ has no branch cuts on 
${\cal  C}_1, \ldots ,{\cal C}_{N-i}$, while $H_{pq}(r)$ has no 
branch cuts on 
${\cal  C}_0, {\cal C}_{-1}, \ldots ,{\cal C}_{1-i}$. We could then  
derive (\ref{myconj1}) by a Wiener-Hopf factorization, taking the 
contour ${\cal C}$ to surround the cuts
${\cal  C}_0, {\cal C}_{-1}, \ldots ,{\cal C}_{1-i}$.

For other values of $i,r$ awkward factors such as $1/(1-\omega)$
occur that multiply the $1/(1-\omega^{j-1}t_p)$ terms, and the 
conditions to use (\ref{Veqns}) in  (\ref{defJr}) are no longer 
satisfied   because $n=-r$ is not in  $\zeta_{k\ell}$. We have 
observed no simple patterns for these cases.



\vspace{-0.4cm}

\section{Summary.}

We have derived the long-conjectured result (\ref{conj}), (\ref{conj2})
for the order parameters of the ferromagnetic chiral Potts model. The 
method does depend on the analyticity assumptions of section 7, but
in this respect it is no different from the standard derivation
of the chiral Potts model free energy. In both calculations one argues
that a certain ``$\tau_2 (t_p)$'' type-function is analytic
except for a single branch cut, and obtains a rule relating the two
values of the function on either side of that 
cut.\cite{RJB90,RJB91,RJB03}

In one respect this order parameter calculation is easier than
that of the free energy. Although we have performed a Wiener-Hopf
factorization to obtain the function $S(t_p)$ of section 8, we did 
not need to. As we remarked at the end of that section, it is 
sufficient to calculate the elementary function $L_r(t_p)$.

It seems that it is this function that is the desired 
generalization to arbitrary $N$ of 
the simple $N = 2$ function $L_{pq}(r)$ of \cite{RJB98}. 
For $N=2$ the chiral Potts model reduces to the Ising model.
There is a difference property and all functions depend on 
$p$, $q$ only via $k$ and $u_q - u_p$, where $u_p, u_q$ are 
elliptic 
function arguments. For a long time the author thought one had
to keep $p, q$ independent to avoid the trivial situation when
$u_q - u_p$ is merely a constant. Of course this is not so:
for $N = 2$ the restriction (\ref{spcase}) corresponds to 
$u_q = 2 K - u_p$, $K$ being a constant (the elliptic integral).
Hence $u_q - u_p = 2 K - 2 u_p$ and the functions do not 
merely degenerate to constants. It is this `` superintegrable''
$N = 2$ case that we have generalized in this paper.

The actual calculation is not difficult, being given in section 8.
Most of this paper is concerned with presenting arguments for the
analyticity assumptions of section 7.

For $N=3$, low-temperature expansions of $G_{pq}(r)$ were 
developed in \cite{RJB98b}. As we remarked above, these 
were very useful in developing and checking the ideas that led 
to our analyticity assumptions. They agree with the results 
(\ref{Gfinres}), (\ref{Cr}), and with the conjecture 
(\ref{myconj1}).

We have only obtained the generalized order parameter function
$G_{pq}(r)$ when $p, q$ are related by (\ref{genrestr}), with 
$i = 0$ and 1. At the end of the previous section we have also 
conjectured how these results may extend to $i = 2, \ldots, N-1$,
$r = 1, \ldots, N-i$. It would be of some interest to study
the other cases of (\ref{genrestr}), and indeed the general case
of no restriction on $p, q$. This last is not an easy problem:
we intend to comment on the difficulties in a 
subsequent paper.

  
\vspace{-0.4cm}


\end{document}